\documentclass
[superscriptaddress,twocolumn,secnumarabic,amssymb,amsmath,nobibnotes,aps,prd,showpacs,showkeys,nofootinbib]{revtex4}%
\usepackage{graphicx}
\usepackage{epsf}
\usepackage{bm}
\usepackage{amsmath}
\usepackage{amsfonts}
\usepackage{amssymb}
\usepackage{epstopdf}%
\providecommand{\U}[1]{\protect\rule{.1in}{.1in}}

\newcommand{\be}{\begin{equation}}
\newcommand{\ee}{\end{equation}}

\newcommand{\mincir}{\raise
-3.truept\hbox{\rlap{\hbox{$\sim$}}\raise4.truept\hbox{$<$}\ }}
\newcommand{\magcir}{\raise
-3.truept\hbox{\rlap{\hbox{$\sim$}}\raise4.truept\hbox{$>$}\ }}

\usepackage{color}
\begin{document}
\title{A model with interaction of dark components and recent
observational data}

\author{Supriya Pan}
\email{span@iiserkol.ac.in}
\affiliation{Department of Physical Sciences, Indian Institute of Science Education and
Research Kolkata, Mohanpur$-$741246, West Bengal, India}

\author{German S. Sharov}
\email{Sharov.GS@tversu.ru}
\affiliation{Tver State University, 170002, Sadovyj per. 35, Tver, Russia}

\begin{abstract}
In the proposed model with interaction between dark energy and dark matter,
we consider cosmological scenarios with different equations of state ($w_d$) for dark
energy. For both constant and variable equation of state, we analyze solutions for dark
energy and dark matter in seven variants of the model. 
We investigate exact analytic solutions
for $w_d={}$ constant equation of state, and several variants of the model for variable
$w_d$. These scenarios are tested with the current astronomical data from Type Ia
Supernovae, baryon acoustic oscillations, Hubble parameter $H (z)$ and the cosmic microwave background radiation. Finally, we make a statistical comparison 
of our interacting model with  $\Lambda$CDM as well as with some other
well known non-interacting cosmological models.
\end{abstract}

\maketitle
\date{\today}

The current  accelerated expansion of the
 universe ascertained from
 several astronomical sources \cite{Riess1998, Perlmutter1999, Percival2001, Spergel2003, Tegmark2004, Eisenstein2005, Komatsu2011} is one of the appealing fields of research in modern
cosmology. To explain such accelerated expansion, either one needs some hypothetical
dark energy fluid in the framework of general relativity or claims direct modifications
into the gravitational sector leading to several modified gravity theories.  However, a
large number of observational data favor the existence of $\Lambda$CDM-cosmology where
the cosmological constant fluid acts as a hypothetical dark energy fluid occupying
almost two third of the total energy density of the universe and the cold
dark matter (CDM) fluid is responsible for the structure formation of the
universe. Nevertheless, the cosmological constant problem \cite{Weinberg1989} and the
cosmic coincidence problem \cite{Steinhardt2003} associated with the
$\Lambda$CDM cosmology lead to several alternative descriptions aiming to decode the
dynamics of the universe, such as the dynamical dark energy models, modified gravity
models, and some others. For a comprehensive discussions on such models and their
effects on the dynamics of the universe, we refer \cite{AT2010}.

Amongst several dark energy and modified gravity models, interacting dark energy models
have gained significant attention due to having a reasonable explanation to the cosmic
coincidence problem. A number of investigations in this direction have been performed
during last couple of years \cite{Wetterich1995, Amendola2000, BC2000,
ZPC2001, AQ2003, HPZ2004, CJPZ2003, HW2008, 
Quartin2008, Maartens2008, Maartens2009, Chimento2010, Andronikos2014, NB2014, Weiqiang1, Weiqiang2, 
Weiqiang3,Sola-et-al-2016}. Recently, a series of
analysis shows that the current observational data prefer a nonzero interaction in the
dark sector \cite{Sola-et-al-2016, Salvatelli2014, NPS2016, KN2016, KN2017, Morrice2017}. On the
other hand, the dark sector's interaction may alleviate the current tension on
local value of the Hubble constant and on $\sigma_8$ as well
\cite{KN2016, PT2016, XW2016, KN2017, AM2017}. 
Further, it has been discussed
that some appropriate choice of an interaction between the dark components can influence
effectively on the perturbation analysis which results in quite notable differences in
the lowest multipoles of the cosmic microwave background spectrum \cite{Zimdahl2005, Wang-et-al-2007}. And sometimes it is possible to address the phantom universe \cite{Caldwell2002,
Planck2014, Planck2015, Rest2014, CH2014, SH2014} by a nonzero coupling in the dark sector \cite{Mohseni2007, PC2014, NPS2016}. For a review
on interacting dark energy and its several consequences see  \cite{Bolotin2015, Wang2016}. Therefore, based on the above arguments it is evident that the possibility
of mutual interaction in the dark sectors may serve as an alternative description to
understand the dynamical evolution of the universe.

In the present work we thus consider the scenario of the universe where DE is
interacting with CDM through a nongravitational interaction $Q$ which in general is an
arbitrary function of the cosmological variables. We consider a particular model
of $Q$ \cite{PBC2015} which is the linear combination of DE and
CDM. We study this interaction model both for constant and variable equation
of state in DE and provide an updated constraints on 
the model parameters using the latest observational
data from Union 2.1, Hubble parameter measurements, baryon acoustic oscillation (BAO)
data and the cosmic microwave background radiation (CMB). Finally, we make a
statistical comparison of the present interacting model with $\Lambda$CDM as well as other known 
non-interacting cosmological models (Sect.~\ref{comparison}).

The paper is organized as follows: In section \ref{dynamics}, we describe the basic
equations of interacting dynamics in the non-flat
FLRW model. For the linear interaction
between DE and CDM, section \ref{analytic} presents the analytic solutions for
$w_d={}$const, in the EoS of DE and 5
 variants of the model for variable $w_d$. In section \ref{data-analysis}, we shortly
discuss different joint data analysis techniques and the results of their application to
the considered several variants of our model. In section \ref{comparison}, we compare
this model with other popular models in describing the same observational data. Finally,
in section \ref{conclu}, we summarize the results of the work.
Let us note that any subindex ``$0$'' attached to any quantity refers its value at present time.

\section{Interacting dynamics in FLRW universe}
\label{dynamics}

Let us consider the homonegeous and isotropic universe characterized by the
Friedmann-Lemm\^itre-Robertson-Walker (FLRW) line element
  $${\rm d}{\rm s}^2= -{\rm d}t^2+
a^2 (t) \left[{\rm d}r^2/(1-kr^2)+ r^2 ({\rm d} \theta^2+ \sin^2 \theta\, {\rm d}
\phi^2)\right],$$
 where $a(t)$ is the scale factor of the universe and $k$ is the
spatial curvature which represents a flat, open and closed universe respectively for $k=
0, -1$, and $1$. In such a background, the first Friedmann equation can be written as
 \begin{eqnarray}
H^2+ \frac{k}{a^2}=\frac{8\pi G}{3}\rho,\label{friedmann1}
\end{eqnarray}
where $H= \dot{a}/a$, is the Hubble rate of the FLRW universe; $\rho$ is the total
energy density of the universe which is the mixture of baryons, cold dark matter and
dark energy, i.e. $\rho= \rho_b + \rho_{dm}+ \rho_{d}$, where $\rho_b$, $\rho_{dm}$,
$\rho_d$ are respectively the energy densities of baryons, CDM and DE. We further 
assume that CDM and DE are interacting with each other while baryons do 
not take part in the interation. The energy
conservation equation for the total fluid follows
\begin{eqnarray}
\dot{\rho}+ 3 \frac{\dot{a}}{a} (p+\rho)= 0.\label{continuity}
\end{eqnarray}
Since only CDM and DE interact with each other but baryons do not interact, thus,
the evolution for baryons follows $\dot{\rho}_b+ 3 H \rho_b = 0$ $\Longrightarrow$ $\rho_b =
\rho_{b0}\, a^{-3}$, while the evolution equations for CDM and DE read
\begin{eqnarray}
\dot{\rho}_{dm}+ 3 H \rho_{dm}&=& Q,\label{conservation1}\\
\dot{\rho}_d+ 3 H (1+ w_d) \rho_d&=& -Q,\label{conservation2}
\end{eqnarray}
where $Q$ is the interaction function between the dark sectors. 
Physically, the interaction is charaterized by some energy 
flow between the sectors interacting with each other. A positive interaction 
(i.e. $Q> 0$) implies the flow of energy
from DE to CDM while its negative value denotes the energy 
flow in the opposite direction. 
Now, introducing the total energy density of  CDM and DE as
$\rho_T= \rho_{dm}+ \rho_d$, it is easy to see that the combination of eqns.
(\ref{conservation1}) and (\ref{conservation2}) turns into
\begin{align}
\dot{\rho}_T+ 3 \frac{\dot{a}}{a} (p_T+\rho_T) &= 0.\label{continuity-total}
\end{align}
Now, using Eq.~(\ref{continuity-total}), one can express $\rho_d$ and $\rho_{dm}$ as follows:
\begin{eqnarray}
 \rho_d &=&-\frac{\rho_T+ \rho^\prime_T}{w_d},\label{de}\\ 
 \rho_{dm} &=&\frac{\rho^\prime_T+(1+ w_d)\rho_T}{w_d}.\label{dm}
\end{eqnarray}
Here primes denote derivatives 
with respect to the variable $x= 3 \ln (a/a_0) = 3 \ln a$ 
(We set $a_0= 1$ as an usual practice and there is no loss of generality). 
Thus, once $\rho_T$ is determined,
the evolution equations for CDM and DE can be understood. However, in the present study we shall concentrate on an interaction function which is the linear combination of the energy densities of CDM and DE. In what follows in the next section we discuss the interacting scenarios both for constant and dynamical equation of state in DE.

 \section{Variants of the model}
\label{analytic}

We introduce the following interaction \cite{Quartin2008, PBC2015}
\begin{equation}
Q= 3\lambda_m H \rho_{dm} +3\lambda_d H \rho_d,\label{interaction1}
\end{equation}
where $\lambda_m$, $\lambda_d$ are the coupling parameters that denote the strength (with their magnitudes) and the direction of energy flow (with their signs) between the interacting sectors.
 Due to the expression (\ref{interaction1}) the conservation equations
(\ref{conservation1}) and (\ref{conservation2}) are modified, and finally, we get the
following second order differential equation:
\begin{eqnarray}
\rho^{\prime\prime}_T &+& \left(2+ w_d+ \lambda_d- \lambda_m- \frac{w^\prime _d}{w_d}
\right) \rho^\prime_T \nonumber  \\
 &+& \left[(1+ w_d)(1- \lambda_m)+ \lambda_d-
\frac{w^\prime _d}{w_d} \right] \rho_T= 0, \label{diffeqn}
\end{eqnarray}
which is the master equation to determine the evolution of CDM and DE. Let us 
 proceed with two different possibilities with the equation of state in DE,
namely when it is either constant or dynamical with the cosmic evolution.

\subsection{The case for constant EoS in DE}

If $w_d={}$ constant, the solution of the differential equation (\ref{diffeqn}) becomes
\cite{Chimento2010}
\begin{equation}
\rho_T= \rho_1 a^{3 m_{+}} + \rho_2 a^{3 m_{-}},\label{energy-const}
\end{equation}
where $\rho_1$, $\rho_2$ are integration constants, $m_{+}$, $m_{-}$ are 
\begin{equation}
m_{\pm}= \frac{\lambda_m-w_d-\lambda_d-2\pm \sqrt{(\lambda_m+
w_d+\lambda_d)^2-4\lambda_m \lambda_d}}{2}.\nonumber
\end{equation}
One can see that for this case, the Hubble function (\ref{F1}) takes an analytic form leading to

\begin{eqnarray*}
H^2 =  \frac{8\pi G}{3} \Bigl[ \rho_{b0} a^{-3} + \rho_1 a^{3m_{+}}+ \rho_2 a^{3m_{-}}  \Bigr] - \frac{k}{a^2}
\end{eqnarray*}

Now, using (\ref{energy-const}), we have the explicit analytic solutions for dust and
dark energy as follows:
\begin{eqnarray}
\rho_{dm}&=& \rho_1 \frac{w_d+1+ m_{+}}{w_d}\, a^{3m_{+}}+ \rho_2 \frac{w_d+1+ m_{-}}{w_d}\, a^{3 m_{-}},\nonumber \label{dust-analytic}\\
\rho_d&=&-\frac{\rho_1 (1+ m_{+})\, a^{3 m_{+}}+ \rho_2 (1+ m_{-})\, a^{3
m_{-}}}{w_d}.\nonumber \label{DE-analytic}
\end{eqnarray}

We mention that in Ref. \cite{PBC2015} the analytic
solution for this particular linear interaction was discussed 
assuming that the magnitudes of both coupling parameters are very small, that means, the
product $\lambda_m \lambda_d$ was excluded and the cosmlogical scenario 
wer analyzed for the solution with $m_{+}= -(1- \lambda_m)$,
$m_{-}= -(1+\lambda_d+ w_d)$. Certainly, a detailed analysis with no such restriction 
is worth investigating. Moreover, the analysis of this model was performed with $194$ Supernovae Type Ia data from \cite{Tonry2003, Barris2004} which needs to be updated with the latest observational data.
Thus, in comparison with  the previous study,
 the present one has two fold importance:
(i) the solution (\ref{energy-const}) for general ($m_{+}, m_{-}$) completes the study
without any information loss, and (ii) here we employ the current observational data
which provide
 better observational constraints on all model parameters. Thus, under
(i) and (ii), the present analytic interacting dark energy model could produce some
interesting information about this interacting dark energy-dark matter model while
constraining it by recent observational data sets.

Further, the usual density parameters for dark matter ($\Omega_{dm0}$) 
and dark energy ($\Omega_{d0}$) in terms of the density
parameters for the equivalent two fluids $\Omega_1$ and $\Omega_2$ are given by
\begin{eqnarray}
 \Omega_{dm0}&=& \Omega_1\frac{w_d+1+ m_{+}}{w_d}+ \Omega_2 \frac{w_d+1+m_{-}}{w_d},\label{Omega_matter} \\
\Omega_{d0}&=& -\frac{\Omega_1 (1+ m_{+})+ \Omega_2
(1+ m_{-})}{w_d},\label{Omega_DE}
\end{eqnarray}
where $\Omega_{i}= 8\pi G\rho_{i}/3 H_0^2$.
The values $\Omega_1$, $\Omega_2$ can be expressed by using the above two equations
(\ref{Omega_matter}), (\ref{Omega_DE}), their consequence  $\Omega_{dm0}+ \Omega_{d0}=
\Omega_1+ \Omega_2$ and  the equality
 $$ 
\Omega_{dm0}+ \Omega_{d0}+\Omega_{b0}+\Omega_{k}=\Omega_{1}+
\Omega_{2}+\Omega_{b0}+\Omega_{k}= 1,
 $$ 
results in from Eq.~(\ref{friedmann1}) at the present time $t=t_0$. Here
$\Omega_{b0}=\Omega_b(t_0)$, $\Omega_{k}= -k/(a_0H_0)^2$. In particular,
\begin{equation}
\Omega_1=\frac{w_d\Omega_{dm0}-(1+w_d+ m_{-})(1-\Omega_{b0}-\Omega_{k})}{m_{+}-m_{-}}.
\label{Omega1}
\end{equation}
Also, we note that the total density parameter for matter is, 
$\Omega_{m0}= \Omega_{dm0}+ \Omega_{b0}$.
In Sect.~\ref{data-analysis} (see Fig.~\ref{F1}) we investigate how solutions
(\ref{energy-const}) 
describe the observational data for Type Ia supernovae, baryon acoustic oscillations,
 for the Hubble parameter $H(z)$ and CMB.

\subsection{Variable EoS in DE} 
\label{variable}

In this section we focus on the interacting models 
where the EoS in DE, $w_d$, is dynamical. 
There are several interacting dark energy models with possibility of variable EoS in DE,
where reasonable attention has been paid to observational data. In Ref.~\cite{WW2014}, the authors investigated an interacting scenario for $Q= 3 H \lambda_m \rho_m$
with Chevallier-Polarski-Linder (CPL) parametrization \cite{CP2001, Linder2003} as the equation of state in DE. Also, in Ref. \cite{HW2008} the authors
studied the present linear interaction (\ref{interaction1}) with CPL parametrization but
with very old data (182 Gold Type Ia Supernoave data \cite{Riess-Gold}). Thus,
considering the linear interaction (\ref{interaction1}) in our discussion, we aim to
investigate the interacting dynamics between CDM an DE with some new variable equations
of state in $w_d$ including CPL \cite{CP2001, Linder2003} and linear
parametrization \cite{Astier2001, CH1999, WA2002} by Union
2.1 compilation \cite{Suzuki2012} along with Hubble parameter measurements,  baryon
acoustic oscillation  and CMB data.

Let us first begin our analysis with the
following generalized ansatz
\begin{equation}
\frac{w^{\prime}_d}{w_d}= \alpha\, w_d +\beta,
 \label{GA}
\end{equation}
where $\alpha$, $\beta$ are real numbers. We note that the EoS (\ref{GA}) is the 
generalized version of the variable EoS of DE presented in 
\cite{PBC2015}.
The solution of Eq.~(\ref{GA}) is
\begin{equation}
w_d= \left[ \left(\frac{1}{w_{d0}} + \frac{\alpha}{\beta}\right) a^{-3\beta} -
\frac{\alpha}{\beta} \right]^{-1}.
 \label{solution-GA}
\end{equation}
In particular, we consider the  following partial cases
\begin{eqnarray}
\mbox{Ansatz I:}\;~~~~\alpha&=&0,\qquad w_d=w_{d0} a^{3\beta}; \label{sol1-GA}\\
\mbox{Ansatz II:}\;~~~~\beta&=&0,\;\;w_d=\frac{w_{d0}}{(1-3 \alpha w_{d0} \ln a)}.
\label{sol2-GA}
\end{eqnarray}
We also consider separately the following ansatz:
\begin{equation}
\mbox{Ansatz III:}~\qquad \alpha=1, \label{Ans3}\\
\end{equation}
which is attractive, because in this case under the condition $\lambda_m= 0$ coefficients
of equation (\ref{diffeqn}) become constant, and its general solution has the simple
form \cite{PBC2015}
\begin{equation}
\rho_T= \tilde{\rho}_1 a^{-3}+ \tilde{\rho}_2 a^{3 (n-1)},\label{variable-total-Energy}
\end{equation}
where $n=\beta-\lambda_d={}$const,  and $\tilde{\rho}_1> 0$, $\tilde{\rho}_2> 0$ are
integration constants.

Moreover, we also consider two more interacting scenarios 
when the EoS of DE obeys the Chevallier-Polarski-Linder (CPL)
parametrization \cite{CP2001, Linder2003}
\begin{equation}
\mbox{Ansatz IV:} \qquad \qquad \qquad w_d (z)= w_{d0}+ w_1\frac{ z}{1+z},
 \label{Ans4}
\end{equation}
and the linear parametrization \cite{Astier2001, CH1999, WA2002}
\begin{equation}
\mbox{Ansatz V:} \qquad \qquad \qquad w_d (z)= w_{d0}+ w_1 z.
 \label{Ans5}\end{equation}
Here in both (\ref{Ans4}), (\ref{Ans5}), $w_{d0}$, and $w_1 = dw_d (z)/dz$ at $z= 0$ 
are two free parameters to be constrained by the observational data. The
dependencies of $w_d (z)$ in (\ref{Ans4}) and (\ref{Ans5}) are alternative to
(\ref{solution-GA}).

\section{Joint Analysis}
\label{data-analysis}

In order to constrain the proposed models with recent observational data,
we use $N_{SN}=580$ data points  for Type Ia supernovae from Union 2.1 \cite{Suzuki2012}, $N_H=39$ observed Hubble data points \cite{Simon2005, Stern2010,%
Moresco2012, Zhang2014, Moresco2015, Moresco2016,
GCH2009, Blake2012, Busca2013, CW2013, Chuang2013, Anderson2014a,%
Anderson2014b, Oka2014, Font-Ribera2014, Delubac2015} and $N_{BAO}=17$ baryon acoustic
oscillation data \cite{GCH2009, Blake2012, Busca2013, CW2013, Chuang2013, Anderson2014a, Anderson2014b, Oka2014, Font-Ribera2014, Delubac2015, Percival2010, Kazin2010, Beutler2011, Blake2011, Padmanabhan2012, Seo2012, Kazin2014, Ross2015, Hinshaw2013}, 
and finally the cosmic microwave background radiation (CMB) 
in the form \cite{HWW2015}.

Our analysis follows the likelihood $\mathcal{L} \propto
\exp(-\chi^2/2)$ where $\chi^2 = \sum_{i} \chi^2_{i}$ ($i$ runs over the all data sets 
employed in the analysis). We calculate the best-fitted values of the free model parameters with their
corresponding uncertainties
from the minimization of the $\chi^2$ function. We use two different combined analysis
with the likelihoods $\mathcal{L}_\Sigma\propto
\exp(-\chi^2_{\Sigma}/2)$, $\mathcal{L}_{tot}\propto \exp(-\chi^2_{tot}/2)$,
where
\begin{eqnarray}
\chi^2_{\Sigma}&=& \chi^2_{SN}+ \chi^2_{H}+ \chi^2_{BAO},\label{total-chi21}\\
\chi^2_{tot}&=& \chi^2_{SN}+ \chi^2_{H}+ \chi^2_{BAO}+ \chi^2_{CMB}.\label{total-chi2}
\end{eqnarray}
In the next subsections, we shall shortly describe different data sets and the corresponding
$\chi^2$ functions.

\subsection{Union 2.1 data points}

Type Ia Supernovae  are the first indication for existence of some dark energy in our
Universe \cite{Riess1998, Perlmutter1999}. The observable quantities from a Type Ia
supernova (SN Ia) are its redshift $z$ and its apparent magnitude $m_{obs}$, but in the
survey \cite{Suzuki2012} values $m_{obs}$ are recalculated into distance modulus
 \begin{equation}
  \mu_{obs}=m_{obs}(z)-M+\bar{\alpha} x_1-\bar{\beta} c+\bar{\delta} P.
  \label{muobs} \end{equation}
Here, additive terms include the SN Ia absolute magnitude $M$  and corrections connected
with deviations from mean values of lightcurve shape ($x_1$), SN Ia color ($c$) and mass
of a host galaxy (the factor $P$). The parameters $M$, $\bar{\alpha}$, $\bar{\beta}$ and
$\bar{\delta}$ are considered in Ref. \cite{Suzuki2012} as nuisance parameters,
{and} they are fitted simultaneously with $H_0$ and other cosmological
parameters in the flat $\Lambda$CDM model. This approach is usual in SN Ia analysis
\cite{NP2005, Conley2011, Ruiz2012}. So values (\ref{muobs}) in Ref. \cite{Suzuki2012} may
have a model dependent additive term (a systematic error) with concealed dependence on
$H_0$ and other model parameters.

We have to keep in mind this fact, when we compare the observable values (\ref{muobs}) from
Ref. \cite{Suzuki2012} with theoretical values of distance modulus, corresponding to
redshift $z$:
\begin{equation}
\mu_{th}(z)= 5 \log_{10} \left(\frac{D_L (z)}{10\mbox{pc}}\right)
=5\log_{10}\frac{H_0D_L}c+\mu_0. 
 \label{mu}
\end{equation}
Here, $\mu_0=42{.}384-5\log_{10}h$, $D_L (z)$ is the luminosity distance \cite{Riess1998, NP2005}
 \begin{equation}
 D_L(z)=\frac{c\,(1+z)}{H_0}S_k
 \bigg(H_0\int\limits_0^z\frac{d\tilde z}{H(\tilde
 z)}\bigg)  \label{DL} \end{equation}
 with
$$S_k(x)=\left\{\begin{array}{ll} \sinh\big(x\sqrt{\Omega_k}\big)\big/\sqrt{\Omega_k}, &\Omega_k>0,\\
 x, & \Omega_k=0,\\ \sin\big(x\sqrt{|\Omega_k|}\big)\big/\sqrt{|\Omega_k|}, &
 \Omega_k<0.
 \end{array}\right.$$
{The} value $H_0D_L/c$ in Eq.~(\ref{mu}) is the Hubble free luminosity distance (for the
majority of cosmological models) and only the term $\mu_0$ \cite{NP2005} depend on the
Hubble constant $H_0$ or $h=H_0/100$ km\,s${}^{-1}$Mpc${}^{-1}$.

For any cosmological model, we fix its model parameters $\theta_1,\theta_2,\dots$,
calculate functions $a(t)$, $z=a^{-1}-1$,
$H(z)$, the integral (\ref{DL}),
 and hence this model predicts theoretical values $D_L^{th}$  or $\mu_{th}$ for the modulus (\ref{mu}). To compare
these theoretical values with the observational data $z_i$ and $\mu_{obs}(z_i)$
\cite{Suzuki2012} we use the $580\times580$ covariance matrix $C_{SN}$  from
Ref. \cite{Suzuki2012} and the function
\begin{equation}
\tilde\chi^2_{SN}(\theta_1,\dots)= \sum_{i,j=1}^{N_{SN}}
 \Delta\mu_i\big(C_{SN}^{-1}\big)_{ij} \Delta\mu_j,\label{chi2a}
\end{equation}
 where $\Delta\mu_i=\mu_{th}(z_i,\theta_1,\dots)-\mu_{obs}(z_i).$

To exclude the  possible systematic errors in
$\mu_{obs}$ mentioned above, we
follow the marginalization procedure, suggested in Ref. \cite{NP2005}, and consider
below the minimum of the sum (\ref{chi2a}) over $H_0$ (or over $\mu_0$)
  \begin{eqnarray}
\chi^2_{SN}=\min\limits_{\mu_0}\tilde\chi^2_{SN}= \tilde\chi^2_{SN}\Big|_{\mu_0=0}&-&
 \frac{B^2}C, \label{chi2m}\\
B=\sum_{i,j=1}^{N_{SN}}(\Delta\mu_i-\mu_0)\big(C_{SN}^{-1}\big)_{ij},\,&\;&\,
C=\sum_{i,j=1}^{N_{SN}} \big(C_{SN}^{-1}\big)_{ij}. \nonumber
\end{eqnarray}

In this paper, for all models we use the marginalized function (\ref{chi2m}) to describe
the SNe Ia data \cite{Suzuki2012}.

\begin{table*}
\begin{tabular}{||l|l|l|c||l|l|l|c||}   \hline
 $z$  & $H_{obs}(z)$ &$\sigma_H$  & References & $z$ & $H_{obs}(z)$ & $\sigma_H$  & References\\ \hline
 0.070 & 69  & 19.6& \cite{Zhang2014}  & 0.570 & 96.8& 3.4 & \cite{Anderson2014b}\\ \hline
 0.090 & 69  & 12  & \cite{Simon2005}  & 0.593 & 104 & 13  & \cite{Moresco2012} \\ \hline
 0.120 & 68.6& 26.2& \cite{Zhang2014}  & 0.600 & 87.9& 6.1 & \cite{Blake2012}   \\ \hline
 0.170 & 83  & 8   & \cite{Simon2005}  & 0.680 & 92  & 8   & \cite{Moresco2012} \\ \hline
 0.179 & 75  & 4   & \cite{Moresco2012}& 0.730 & 97.3& 7.0 & \cite{Blake2012}  \\ \hline
 0.199 & 75  & 5   & \cite{Moresco2012} & 0.781 & 105 & 12  & \cite{Moresco2012}\\ \hline
 0.200 & 72.9& 29.6& \cite{Zhang2014}  & 0.875 & 125 & 17  & \cite{Moresco2012}\\ \hline
 0.240 &79.69& 2.99& \cite{GCH2009}  & 0.880 & 90  & 40  & \cite{Stern2010} \\ \hline
 0.270 & 77  & 14  & \cite{Simon2005}  & 0.900 & 117 & 23  & \cite{Simon2005}  \\ \hline
 0.280 & 88.8& 36.6& \cite{Zhang2014}  & 1.037 & 154 & 20  & \cite{Moresco2012}\\ \hline
 0.300 & 81.7& 6.22& \cite{Oka2014}    & 1.300 & 168 & 17  & \cite{Simon2005}  \\ \hline
 0.340 & 83.8& 3.66& \cite{GCH2009}  & 1.363 & 160 & 33.6& \cite{Moresco2015}\\ \hline
 0.350 & 82.7& 9.1 & \cite{CW2013}& 1.430 & 177 & 18  & \cite{Simon2005}  \\ \hline
 0.352 & 83  & 14  & \cite{Moresco2012}& 1.530 & 140 & 14  & \cite{Simon2005}  \\ \hline
 0.400 & 95  & 17  & \cite{Simon2005}  & 1.750 & 202 & 40  & \cite{Simon2005}  \\ \hline
 0.429 & 91.8& 5.3 & \cite{Moresco2016}& 1.965 &186.5& 50.4& \cite{Moresco2015}\\ \hline
 0.430 &86.45& 3.97& \cite{GCH2009}  & 2.300 & 224 & 8.6 & \cite{Busca2013}  \\ \hline
 0.440 & 82.6& 7.8 & \cite{Blake2012}  & 2.340 & 222 & 8.5 & \cite{Delubac2015}\\ \hline
 0.480 & 97  & 62  & \cite{Stern2010}  & 2.360 & 226 & 9.3 & \cite{Font-Ribera2014}\\  \hline
 0.570 & 87.6& 7.8 & \cite{Chuang2013} &       &     &     &                 \\  \hline
 \end{tabular}
\caption{Hubble parameter values $H_{obs}$ in km\,s$^{-1}$Mpc$^{-1}$ at different
redshifts $z$ with corresponding errors $\sigma_H$.} \label{H-data}
\end{table*}

\subsection{Hubble parameter data}

The Hubble parameter $H$ at some certain redshift $z$ can be measured from differential
ages of galaxies \cite{Simon2005, Stern2010, Moresco2012, Zhang2014, Moresco2015, Moresco2016} with using
the following formula:
 $$ 
 H (z)= \frac{\dot{a}}{a} = -\frac{1}{1+z}
\frac{dz}{dt}, $$
In addition, estimations of $H(z)$ may be extracted from line-of-sight BAO data
\cite{GCH2009, Blake2012, Busca2013, CW2013, Chuang2013, Anderson2014a, Anderson2014b, Oka2014, Font-Ribera2014, Delubac2015}.

In this analysis we use $N_H=39$ observed Hubble parameter values
\cite{Simon2005, Stern2010, Moresco2012, Zhang2014, Moresco2015, Moresco2016, GCH2009, Blake2012, Busca2013, CW2013, Chuang2013, Anderson2014a, Anderson2014b, Oka2014, Font-Ribera2014, Delubac2015} in the range $0.070 \leq z \leq 2.36$, which are listed in Table (\ref{H-data}). The
corresponding $\chi^2_{H}$ is defined as
\begin{equation}
\chi^2_{H}= \sum_{i=1}^{N_H} \left[\frac{H_{obs}(z_i)-H_{th}(z_i,
\theta_j)}{\sigma_{H,i}}\right]^2.\label{chi-OHD}
\end{equation}

\subsection{BAO data}

\begin{table*}
\begin{tabular}{||l|l|l|l|l|c|l||} \hline
 $z$  & $d_z(z)$ &$\sigma_d$    & ${ A}(z)$ & $\sigma_A$  & References & Survey\\ \hline
 0.106& 0.336  & 0.015 & 0.526& 0.028& \cite{Beutler2011, Hinshaw2013}  & 6dFGS \\ \hline
 0.15 & 0.2232 & 0.0084& -    & -    & \cite{Ross2015} & SDSS DR7  \\ \hline
 0.20 & 0.1905 & 0.0061& 0.488& 0.016& \cite{Percival2010, Blake2011}  & SDSS DR7 \\ \hline
 0.275& 0.1390 & 0.0037& -    & -    & \cite{Percival2010}& SDSS DR7 \\ \hline
 0.278& 0.1394 & 0.0049& -    & -    & \cite{Kazin2010}  &SDSS DR7 \\ \hline
 0.314& 0.1239 & 0.0033& -    & -    & \cite{Blake2011}& SDSS LRG \\ \hline
 0.32 & 0.1181 & 0.0026& -    & -    & \cite{Anderson2014b} &BOSS DR11 \\ \hline
 0.35 & 0.1097 & 0.0036& 0.484& 0.016& \cite{Percival2010, Blake2011} &SDSS DR7 \\ \hline
 0.35 & 0.1126 & 0.0022& -    & -    & \cite{Padmanabhan2012}   &SDSS DR7 \\ \hline
 0.35 & 0.1161 & 0.0146& -    & -    & \cite{CW2013}   &SDSS DR7 \\ \hline
 0.44 & 0.0916 & 0.0071& 0.474& 0.034& \cite{Blake2011} & WiggleZ \\ \hline
 0.57 & 0.0739 & 0.0043& 0.436& 0.017& \cite{Chuang2013}& SDSS DR9 \\ \hline
 0.57 & 0.0726 & 0.0014& -    & -    & \cite{Anderson2014b} & SDSS DR11 \\ \hline
 0.60 & 0.0726 & 0.0034& 0.442& 0.020& \cite{Blake2011} & WiggleZ \\ \hline
 0.73 & 0.0592 & 0.0032& 0.424& 0.021& \cite{Blake2011} &WiggleZ \\ \hline
 2.34 & 0.0320 & 0.0021& -& - & \cite{Delubac2015} & BOSS DR11 \\ \hline
 2.36 & 0.0329 & 0.0017& -& - & \cite{Font-Ribera2014} & BOSS DR11 \\  \hline
 \end{tabular}
\caption{Values of $d_z(z)=r_s(z_d)/D_V(z)$ and $A(z)$ (\ref{dzAz}) with errors and
references} \label{TBAO}
\end{table*}

Observational data, connected with baryon acoustic oscillations (BAO), include the
distance \cite{Eisenstein2005}
$$
 D_V(z)=\bigg[\frac{cz D_L^2(z)}{(1+z)^2H(z)}\bigg]^{1/3},
 $$
and  two measured  values
 \begin{equation}
 d_z(z)= \frac{r_s(z_d)}{D_V(z)},\qquad
  A(z) = \frac{H_0\sqrt{\Omega_{m0}}}{cz}D_V(z).
  \label{dzAz} \end{equation}
Here $r_s(z_d)$ is sound horizon size at the end of the drag era $z_d$.
 In this paper we
use the fitting formula from Ref.~\cite{Aubourg2015}
 \begin{equation}
 r_s(z_d)=\frac{55.154\exp\big[72.3(\Omega_\nu h^2 + 0.0006)^2\big]}
{(\Omega_{m0} h^2)^{0.25351} (\Omega_{b0} h^2)^{0.12807}}\mbox{ Mpc}.
 \label{rsA} \end{equation}
 Here dependence on neutrino contribution $\Omega_\nu$ is negligible for  reasonable values
$\sum m_\nu \le 0.23$ eV \cite{Planck2015} (below we suppose $\sum m_\nu=0.06$ eV \cite{Planck2015, Aubourg2015}).

Calculations with similar observational data and with the function (\ref{rsA}) were made
in Ref. \cite{Sharov2016} for the models: $\Lambda$CDM, with generalized and modified
Chaplygin gas and with quadratic equation of state (described below in
Sect.~\ref{comparison}). The best fitting value of $\Omega_{b0}$ in Eq.~(\ref{rsA})
 \begin{equation}
 \Omega_{b0}=0.044
 \label{Omb} \end{equation}
 was obtained for the $\Lambda$CDM and appeared to be just the same for 3 other models
in Ref. \cite{Sharov2016}. One should note that the value (\ref{Omb}) is connected with the
formula (\ref{rsA}). Calculations in Ref. \cite{Sharov2016} with the more simple fitting
formula  $r_d=(r_d h)_{fid}\cdot h^{-1}$ for all 4 models demonstrated similar
estimations of model parameters, but very weak dependence of them on $\Omega_{b0}$. It
is connected with similarity in properties of dark matter and baryons. Due to this
reason we do not consider $\Omega_{b0}$ as a free model parameter and fix it in the form
(\ref{Omb}) for all models in this paper. The additional reason is necessity to minimize
a number of free model parameters for considered scenarios.

To take into account all available BAO data \cite{GCH2009, Blake2012, Busca2012, CW2013, Chuang2013, Anderson2014a, Anderson2014b, Oka2014, Font-Ribera2014, Delubac2015, Percival2010, Kazin2010, Beutler2011, Blake2011, Padmanabhan2012, Seo2012, Kazin2014, Ross2015, Hinshaw2013}
for parameters (\ref{dzAz}), we consider in this paper $N_B=17$ data points for $d_z(z)$
and 7 data points for $A(z)$ presented in the Table~\ref{TBAO}.

Measurements of $d_z(z)$ and $A(z)$ from Refs.~\cite{Percival2010, Blake2011} in
Table~\ref{TBAO}
 are not independent. So the $\chi^2$ function for
the values (\ref{dzAz}) is
 \begin{equation}
 \chi^2_{BAO}(\theta_j)=(\Delta d)^TC_d^{-1}\Delta d+
(\Delta { A})^TC_A^{-1}\Delta  A,
  \label{chiB} \end{equation}
  where $\Delta d=d_z(z_i)-d_z^{th}$, $\Delta A=A(z_i)-A^{th}$.
 The elements of covariance matrices
$C_d^{-1}=||c^d_{ij}||$, $C_A^{-1}=||c^A_{ij}||$ in Eq.~(\ref{chiB}) are
\cite{Percival2010, Blake2011, Hinshaw2013}:
$$\begin{array}{lll}
c^d_{33}=30124,& c^d_{38}=-17227,&  c^d_{88}=86977, \\
c^d_{1\!11\!1}=24532.1, & c^d_{1\!11\!4}=-25137.7,& c^d_{1\!11\!5}=12099.1,\\
c^d_{1\!41\!4}=134598.4,& c^d_{1\!41\!5}=-64783.9,& c^d_{1\!51\!5}=128837.6;\\
c^A_{1\!11\!1}=1040.3,  & c^A_{1\!11\!4}=-807.5,  & c^A_{1\!11\!5}=336.8, \\
c^A_{1\!41\!4}=3720.3,  & c^A_{1\!41\!5}=-1551.9, & c^A_{1\!51\!5}=2914.9.
 \end{array}$$
 Here $c_{ij}=c_{ji}$, the remaining matrix elements are $c_{ij}=0$, if $i\ne j$,
and $c_{ii}=1/\sigma_i^2$.

\subsection{CMB data}
Cosmological data associated with the cosmic microwave background (CMB) radiation
include parameters at the photon-decoupling epoch $z_*=1089.90 \pm0.30$ \cite{Planck2015}, in particular, the comoving sound horizon $r_s(z_*)$  and the
distance $D_M(z_*)=D_L(z_*)\big/(1+z_*)$ \cite{Aubourg2015, HWW2015}. In this paper we use the CMB parameters in the form \cite{HWW2015}
 $$ 
  \mathbf{x}=\big(R,\ell_A,\omega_b\big)=\bigg(\sqrt{\Omega_m}\frac{H_0D_M(z_*)}c,\,\frac{\pi D_M(z_*)}{r_s(z_*)},\,\Omega_bh^2\bigg).
 $$ 
In the corresponding $\chi^2$ function
  \begin{equation}
\chi^2_{CMB}=\Delta\mathbf{x}\cdot C_{CMB}^{-1}\big(\Delta\mathbf{x}\big)^{T},
 \label{chiCMB} 
 \end{equation}
we use the covariance matrix $C_{CMB}$ and the distance priors
 $$\Delta \mathbf{x}=\mathbf{x}-\big(1.7448,\; 301.46,\; 0.0224\big),$$
from Ref.~\cite{HWW2015}, which were derived from \cite{Planck2015} data with free amplitude of the lensing power spectrum.

\begin{figure*}
\centerline{\includegraphics[scale=0.72,trim=3mm 0mm 2mm -1mm]{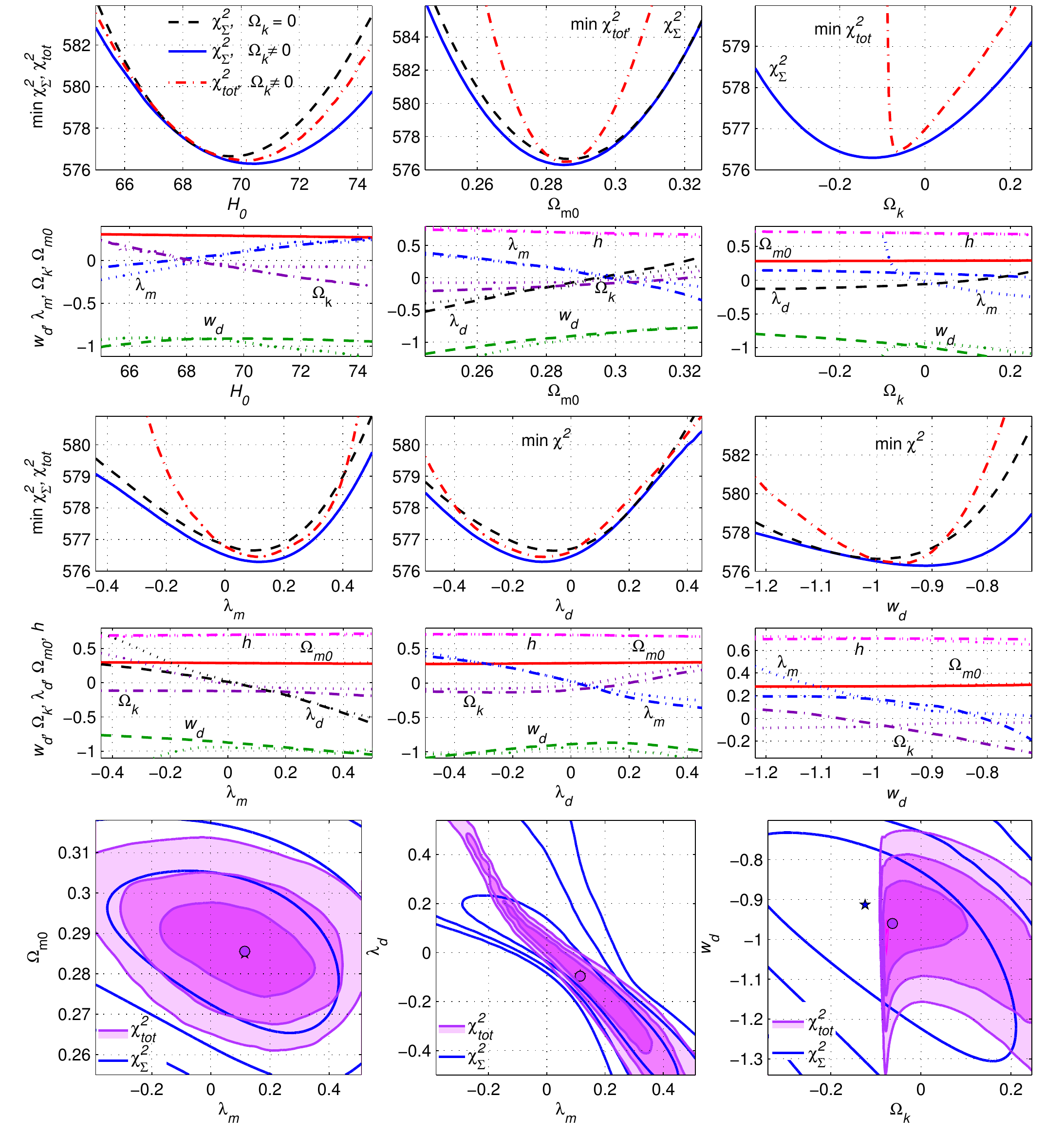}}
  \caption{For the model (\ref{interaction1}) with  $w_d={}$const
in the first and third rows of panels we present dependence of
 $\min\chi^2_\Sigma$ and $\min\chi^2_{tot}$ on $H_0$, $\Omega_{m0}$, $\Omega_k$ , $\lambda_m$, $\lambda_d$ and $w_d$
 and also (in the panels below) the correspondent  dependence for parameters of a minimum point.
In the bottom panels the contour plots in the planes of 2 parameters are drawn at
$1\sigma$, $2\sigma$ and $3\sigma$ confidence levels for $\chi^2_\Sigma$ (blue lines)
and $\chi^2_{tot}$ (filled contours).}
  \label{F1}
\end{figure*}

\subsection{Results for $w_d ={}$ constant.} 
\label{result-1}
We investigated, how the $\chi^2$ functions (\ref{total-chi21}) and
(\ref{total-chi2}) depend on model parameters for different 
variants of the model, considered in
Sect.~\ref{analytic}. For the model with $w_d={}$const
(equivalently, solution (\ref{energy-const})),
the constraints on the model parameters are presented in Table~\ref{Estim} and 
the corresponding plots are shown in Fig.~\ref{F1}.
The first and third rows of panels in Fig.~\ref{F1} illustrate how the minimum of the sums
(\ref{total-chi21}) ($\min\chi^2_\Sigma$) and (\ref{total-chi2}) ($\min\chi^2_{tot}$)
depend on one chosen parameter:  $H_0$, $\Omega_{m0}$, $\lambda_m$, $\lambda_d$,
$\Omega_k$ and $w_d$. Here, for $\chi^2_\Sigma$ we compare two cases: for the model with
6 free parameters including $\Omega_{k}\ne0$ where such dependencies are shown as blue thick
lines, but for the flat case $\Omega_k=0$, the corresponding plots are black dashed
lines. The graphs for $\chi^2_{tot}$ with CMB (red dash-dotted lines) are made for the
general case $\Omega_{k}\ne0$.

In particular, in the top-left panel of Fig.~\ref{F1}, the function $\min\chi^2_{tot}(H_0)$
means $\min\limits_{\Omega_{m0},\Omega_k,w_d,\lambda_m,\lambda_d}\chi^2_{tot}$ (and the
similar minimum for $\chi^2_\Sigma$). The $\chi^2$  absolute minima for these cases are
presented  in Table~\ref{Estim} with optimal values and $1\sigma$ errors of model
parameters. For each considered variant of the model (\ref{interaction1}) the
corresponding line  in Table~\ref{Estim} is obtained from the joint analysis
SNe+$H(z)$+BAO (for $\chi^2_\Sigma$) and the lower case of the line includes the
absolute minimum of $\chi^2_{tot}$ and estimations from the joint analysis SNe+$H(z)$+BAO+CMB.

For example, for the variant $w_d={}$const with $\Omega_k\ne0$, we estimate the Hubble constant
$ H_0=70.40_{-2.13}^{+2.18}$ km\,(s\,Mpc)${}^{-1}$ for $\chi^2_\Sigma$ and
$H_0=70.18_{-1.97}^{+1.77}$ km\,(s\,Mpc)${}^{-1}$ for
 $\chi^2_{tot}$; $1\sigma$ errors are extracted from the
one-dimensional likelihood function ${\cal L}\propto\exp(-\chi^2/2)$.

The similar estimation for $\Omega_{m0}$ is determined by the functions
$\min\chi^2_j(\Omega_{m0})=\min\limits_{H_0,\Omega_k,w_d,\lambda_m,\lambda_d}\chi^2_j$,
if $\Omega_k\ne0$; $j=\Sigma,\,{tot}$. These graphs for $\chi^2_\Sigma$ in the flat and
non-flat cases are rather close and have distinct minimum with small $1\sigma$ deviation
$\Delta\Omega_{m0}\simeq0.013$. In the case $\chi^2_{tot}$ the minimum is the same, but
with smaller $\Delta\Omega_{m0}\simeq0.008$. It is connected with the factor
$\sqrt{\Omega_{m0}}$ in the values $A(z)$ in (\ref{dzAz}) and $R$ in (\ref{chiCMB}), so the
contributions of $\chi^2_{BAO}$ and $\chi^2_{CMB}$ in the sum (\ref{total-chi2}) are very
sensitive to $\Omega_{m0}$ values.

Dependence of $\min\chi^2_{tot}$ on the curvature $\Omega_k$ in the top-right panel of
Fig.~\ref{F1} is strongly asymmetric, unlike the case of $\min\chi^2_\Sigma$. In these
cases we have different $1\sigma$ intervals for $\Omega_k$ (see Table~\ref{Estim}),
but both include values $\Omega_k\simeq0$. Some  asymmetry may be seen for the plots
$\min\chi^2(\lambda_m)$,  $\min\chi^2(\lambda_d)$ and  $\min\chi^2(w_d)$ in the third
row of panels. These calculations result in the estimations of $\lambda_m$, $\lambda_d$
and $w_d$ in Table~\ref{Estim}.

 \begin{table*}
 \begin{tabular}{|l|l|c|c|c|c|c|c|c|}  \hline
Variant  & Data$\!$ &$\min\chi^2$& $H_0$&  $\Omega_{m0}$& $\lambda_m$& $\lambda_d$& $w_{d0}$ & 6th parameter  \\
\hline %
 $w_d={}$const & $\chi^2_\Sigma$ & 576.29 & $70.40_{-2.13}^{+2.18}$ & $0.285\pm0.013$ & $0.115_{-0.265}^{+0.217}$ &
 $-0.093_{-0.259}^{+0.230}$ &   $-0.913_{-0.214}^{+0.132}$ & $\Omega_k=-0.124_{-0.190}^{+0.213}$\rule{0pt}{1.2em}  \\
                & $\chi^2_{tot}$ & 576.45 & $70.18_{-1.97}^{+1.77}$ & $0.285\pm0.008$ & $0.115_{-0.193}^{+0.208}$ &
 $-0.097_{-0.240}^{+0.212}$ &   $-0.955_{-0.113}^{+0.070}$ &  $\Omega_k=-0.064_{-0.019}^{+0.102}$\rule{0pt}{1.2em}  \\
  \hline
  $w_d={}$const & $\chi^2_\Sigma$ & 576.64 & $69.68_{-1.75}^{+1.80}$   & $0.287\pm0.013$ & $0.090_{-0.260}^{+0.213}$ & $-0.059_{-0.280}^{+0.233}$ &
  $-0.994_{-0.157}^{+0.123}$ & $-$\rule{0pt}{1.2em}  \\
  $\;$\&\,$\Omega_k=0$  & $\chi^2_{tot}$ &  576.98 & $69.31_{-1.52}^{+1.67}$   & $0.285\pm0.008$ & $0.039_{-0.104}^{+0.193}$ & $-0.024_{-0.184}^{+0.168}$ &
  $-0.987_{-0.074}^{+0.096}$ & $-$\rule{0pt}{1.2em} \\
 \hline
  Ansatz I  & $\chi^2_\Sigma$& 576.29& $69.55_{-1.73}^{+1.80}$  & $0.288\pm0.013$ & $0.173_{-0.32}^{+0.155}$ & $-0.277_{-0.309}^{+0.407}$ &
  $-0.94_{-0.157}^{+0.137}$\rule{0pt}{1.2em} &  $\beta=-0.25_{-0.54}^{+0.77}$ \\
   $\;(\alpha=0)$  & $\chi^2_{tot}$ &  576.93& $69.86_{-1.68}^{+1.73}$  & $0.286\pm0.008$ & $0.245_{-0.34}^{+0.42}$ & $0.120_{-0.114}^{+0.136}$ &
  $-1.092_{-0.125}^{+0.148}$\rule{0pt}{1.2em} &  $\beta=0.34_{-0.475}^{+0.12}$  \\
   \hline
   Ansatz II  & $\chi^2_\Sigma$ & 576.57& $69.66_{-1.75}^{+1.80}$  & $0.287\pm0.013$ & $0.123_{-0.175}^{+0.18}$ & $-0.112_{-0.26}^{+0.285}$ &
  $-0.988_{-0.135}^{+0.11}$ & $\alpha=0.073_{-0.055}^{+0.062}$\rule{0pt}{1.2em}  \\
   $\;(\beta=0)$  & $\chi^2_{tot}$ & 576.95& $70.22_{-1.72}^{+1.69}$ & $0.288_{-0.008}^{+0.007}$ & $-0.035_{-0.08}^{+0.047}$ & $0.075_{-0.097}^{+0.125}$ &
  $-0.980_{-0.145}^{+0.13}$ & $\alpha=-0.032_{-0.064}^{+0.068}$\rule{0pt}{1.2em}   \\
   \hline
  Ansatz III  & $\chi^2_\Sigma$& 576.64& $69.68_{-1.74}^{+1.80}$  & $0.287\pm0.013$ & $0.098_{-0.24}^{+0.475}$ & $-0.060_{-0.266}^{+0.74}$ &
  $-0.996_{-0.155}^{+0.12}$ & $\beta=0.997_{-0.115}^{+0.243}$\rule{0pt}{1.2em}  \\
   $\;(\alpha=1)$  & $\chi^2_{tot}$ & 577.33& $68.82_{-1.35}^{+1.48}$  & $0.290\pm0.007$ & $0.022_{-0.048}^{+0.093}$ & $-0.548_{-0.42}^{+0.58}$ &
  $-0.96_{-0.142}^{+0.095}$ & $\beta=0.235_{-0.10}^{+0.216}$\rule{0pt}{1.2em}  \\
   \hline
  Ansatz IV  & $\chi^2_\Sigma$& 576.07& $69.23_{-1.86}^{+1.90}$  & $0.292_{-0.015}^{+0.016}$ & $0.237_{-0.250}^{+0.076}$ & $-0.452_{-0.43}^{+0.73}$ &
  $-0.786_{-0.31}^{+0.356}$ & $w_1=-3.14_{-4.72}^{+4.30}$\rule{0pt}{1.2em}  \\
   $\;$Eq. (\ref{Ans4})  & $\chi^2_{tot}$ & 576.90& $69.14_{-1.62}^{+1.74}$  & $0.285\pm0.008$ & $0.016_{-0.038}^{+0.044}$ & $-0.054_{-0.110}^{+0.096}$ &
  $-0.925_{-0.215}^{+0.23}$ & $w_1=-0.68_{-1.14}^{+0.90}$\rule{0pt}{1.2em}  \\
  \hline
  Ansatz V  & $\chi^2_\Sigma$& 575.97& $69.32_{-1.74}^{+1.84}$ & $0.292\pm0.015$ & $0.220_{-0.245}^{+0.062}$ & $-0.497_{-0.386}^{+0.64}$ &
  $-0.810_{-0.27}^{+0.338}$ & $w_1=-2.68_{-4.03}^{+3.75}$\rule{0pt}{1.2em} \\
   $\;$Eq.(\ref{Ans5})  & $\chi^2_{tot}$ &576.92& $69.18_{-1.65}^{+1.70}$ & $0.292\pm0.008$ & $0.023_{-0.042}^{+0.030}$ & $-0.092_{-0.120}^{+0.084}$ &
  $-0.822_{-0.255}^{+0.192}$ & $w_1=-0.43_{-1.57}^{+0.65}$\rule{0pt}{1.2em}  \\
  \hline
   \end{tabular}
 \caption{{Variants of the model (\ref{interaction1}) and $1\sigma$ estimates of the model parameters
 using the joint analysis 
SNe+$H(z)$+BAO (the upper case on all lines) and SNe+$H(z)$+BAO+CMB (the lower case on all lines).}}
 \label{Estim}
 \end{table*}

The panels in the second and forth rows of Fig.~\ref{F1} correspond to the above panels
and present dependencies of coordinates of minima points (optimal values of parameters)
on $H_0,\dots$, $\Omega_k$ for the function $\chi^2_\Sigma$ (if $\Omega_{k}\ne0$) as
thick lines and for $\chi^2_{tot}$ as dots. One can see that optimal values of
$\lambda_m$ and $\lambda_d$ have distinct negative correlation (observed explicitly in
the middle bottom panel), optimal values of $\Omega_{m0}$ and $h=H_0/100$ depend on
other parameters rather weakly.

In 3 bottom panels of Fig.~\ref{F1} we present the
$1\sigma$ (68.27\%), $2\sigma$ (95.45\%) and $3\sigma$ (99.73\%) contour plots  for the
functions $\min\chi^2_j(p_1,p_2)$ in the planes of two parameters. The minimum is
calculated over the remaining 4  parameters.  The mentioned level lines are shown for
the $\chi^2_\Sigma$ function  (blue curves) and for $\chi^2_{tot}$  as filled contours.
The circles and stars in the plots mark minimum points obtained respectively for $\chi^2_{tot}$ and
$\chi^2_\Sigma$.


\subsection{Results for $w_d \neq $ constant. }
\label{result-2}

For the variable EoS in DE, presented in section \ref{variable}, 
we summarize their observational constraints in Table \ref{Estim}
for both SNe+$H(z)$+BAO and SNe+$H(z)$+BAO+CMB. We consider 
several possibilities for variable $w_d$. 
The first one is very general given in Eq.~(\ref{GA})
and it provides three distinct possibilities in 
equations (\ref{sol1-GA}), (\ref{sol2-GA}) and (\ref{Ans3}) while 
additionally we consider CPL (\ref{Ans4}) and linear 
parametrizations (\ref{Ans5}). 
The general ansatz (\ref{GA}) with its solution
(\ref{solution-GA}) gives not only new possibilities, but also additional problems of
the following two types: (i) two extra model parameters (3 parameters $w_{d0}$, $\alpha$
and $\beta$ instead of one $w_d$); (ii) singularities in the past, which appear in
different scenarios of the class (\ref{solution-GA}).

These singularities connected with bad behavior of densities $\rho_{dm}$, $\rho_d$ or
their sum $\rho_T$ at a moment $t_s$ in the past, when the scale factor remains finite
and nonzero  $\big(a(t_s)\ne0\big)$, they may be classified into the following three
types:
 \begin{eqnarray}
a) &\;&\lim\limits_{t\to t_s}\rho_T=\infty;\nonumber\\
b) &\;&\rho_{dm}<0, \mbox{ \ if \ } t< t_s; \label{singul}\\
c) &\;&\rho_d<0, \mbox{\,  \ if \ } t< t_s.\nonumber
 \end{eqnarray}

\begin{figure}
\centerline{\includegraphics[width=0.45\textwidth]{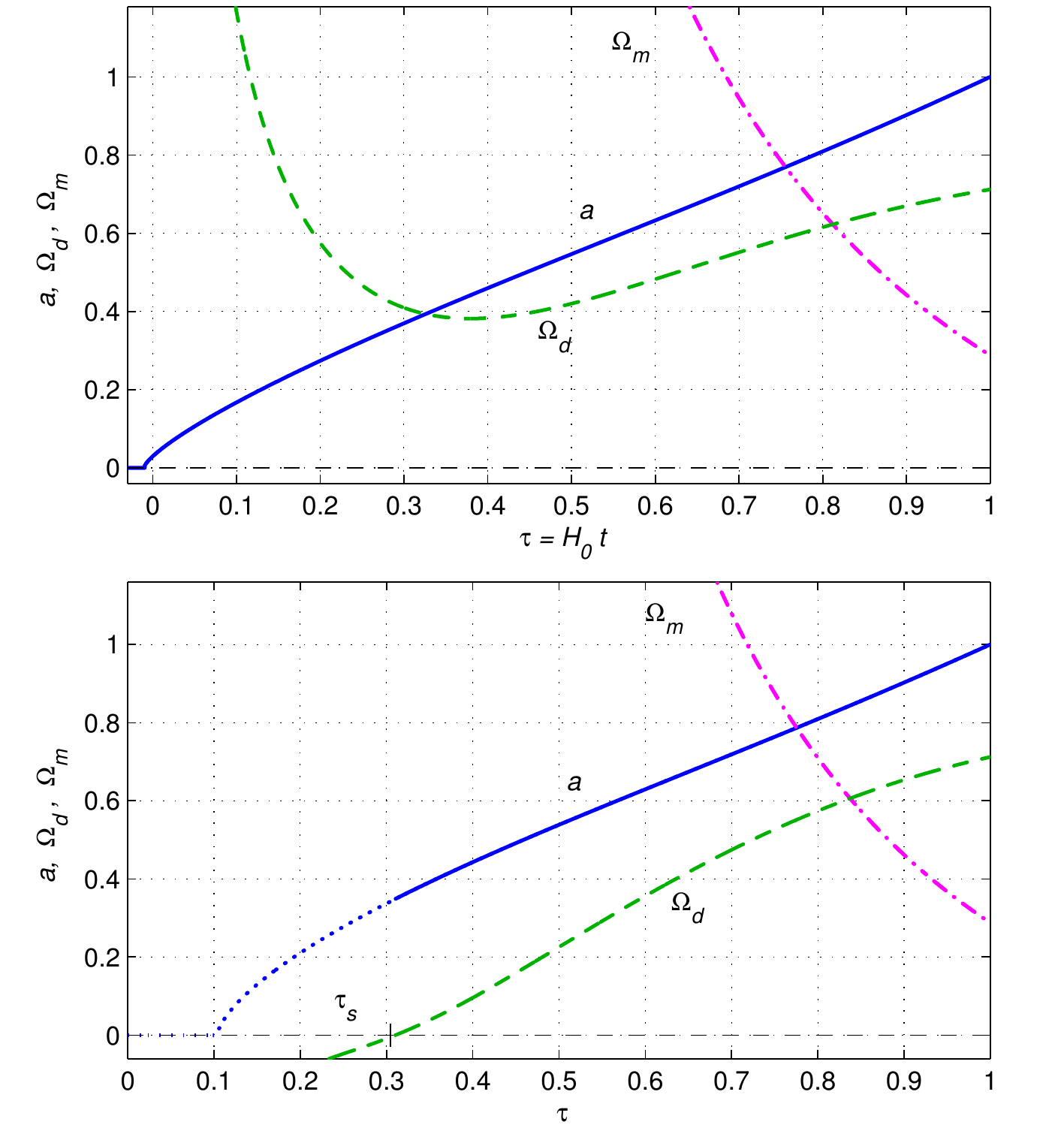}}
 \caption{ Evolution of the
scale factor $a(\tau)$ and  densities $\Omega_m (\tau)$,  $\Omega_d (\tau)$ for Ansatz I
(\ref{sol1-GA}) is shown for the regular solution (top) with optimal parameters from
Table~\ref{Estim} and for the singular solution (bottom) with  type (c1) singularity
(\ref{singul}) (here $\lambda_m= -0.01$, other parameters are the same).}
  \label{Fsing}
\end{figure}

These cases resemble classification of singularities in Refs. \cite{NOT2005, NO2010,%
Bamba-et-al-2012}.
For singularities (\ref{singul}) of the type (c) DE pressure $p_d(t_s)$ remains finite
at the moment $t_s$, whereas $\rho_d(t_s)=0$;  they may be also divided into class (c1)
with $p_d(t_s)=0$ and finite $w_d=p_d/\rho_d$ and class (c2) with $p_d(t_s)\ne0$, where
$w_d$  tends to infinity if $t\to t_s$. Possible singularities (\ref{singul}) compel us
to be especially careful, when we calculate numerically parameters of effective
scenarios in this model. We should exclude domains in parameter space with singular
behavior of physical densities irrespective of type (\ref{singul}). The example of
singular solution with the type (c1) singularity (\ref{singul}) is shown
 in Fig.~\ref{Fsing}.

In  Fig.~\ref{Fsing} we compare the regular solution for Ansatz I (\ref{sol1-GA}) (the
top panel) with the singular solution in the bottom panel. One can see how the scale
factor $a$ (blue solid lines) and densities
 $\Omega_m= \rho_m/\rho_{cr}$ (magenta dash-dotted lines) and $\Omega_d(\tau)= \rho_d/\rho_{cr}$ (green dashed lines)
depend on dimensionless time $\tau=H_0t$. Here $\rho_{cr}= 3 H_0^2/ 8 \pi G$ is the
critical density of the universe. The model parameters are taken with their optimal
values from Table~\ref{Estim} for $\chi^2_\Sigma$, but for the type (c1) singularity
(the bottom panel) with difference only in one value: $\lambda_m=-0.01$. In this
singular case the DE density $\Omega_d$ becomes negative at $\tau<\tau_s$.

We  mentioned above, that the number $N_p$ of  model parameters for scenarios with
variable $w_d$ satisfying Eq.~(\ref{GA}) is too large, it is disadvantage in competition
with other models in accordance with information criteria \cite{Akaike1974, SKK2006, SHL2012}.
So we have to exclude  non-flat scenarios and fix in this section $\Omega_k=0$.  But
even for the flat case we have $N_p=7$ parameters: $H_0$, $\Omega_{m0}$, $\lambda_m$,
$\lambda_d$, $w_{d0}$, $\alpha$, $\beta$.

An attempt to  exclude $\lambda_m$ appeared to be unsuccessful: though in the case
$\lambda_m=0$, we can avoid some  singularities (\ref{singul}) and 
instabilities in perturbations \cite{HWA2008, VMM2008}, but the best value of the
function (\ref{total-chi2}) $\min\chi^2_\Sigma\simeq576.74$ is worth, than in the case
$w_d={}$const  (\ref{energy-const}) (but $\lambda_m\ne0$). So we have to fix other
parameters. First, we consider the case  $\alpha=0$ (\ref{sol1-GA}), denoted in
Sect.~\ref{variable} as Ansatz I.

For Ansatz I ($\alpha=0$) we have no acceptable analytic solution of
Eq.~(\ref{diffeqn}), so we investigate numerical solutions of the system
(\ref{friedmann1}), (\ref{conservation1}), (\ref{conservation2}), (\ref{interaction1}),
(\ref{sol1-GA}) with natural initial conditions $\rho_m\big|_{t=t_0}=\rho_{m0}$,
$\rho_d\big|_{t=t_0}=\rho_{d0}$ at the present day and integration ``into the past''.
For Ansatz I we can reach the best values  $\min\chi^2_\Sigma\simeq576.29$ (this
solution is shown in Fig.~\ref{Fsing} in the top panel) and
$\min\chi^2_{tot}\simeq576.93$ . The corresponding values of model parameters are
tabulated in the ``Ansatz I'' line of Table~\ref{Estim}.

\begin{figure*}
\centerline{\includegraphics[width=0.7\textwidth]{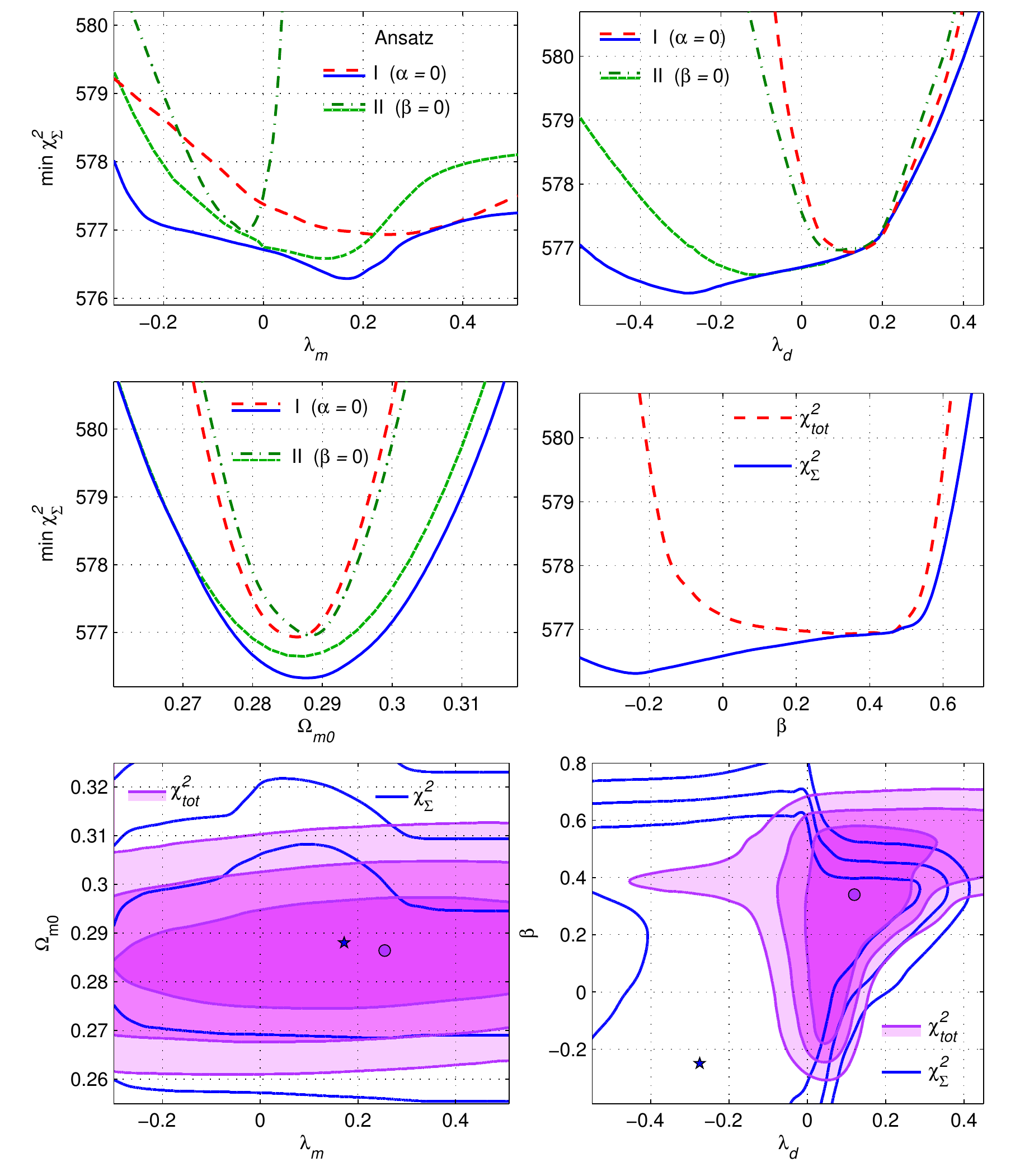}}
  \caption{For Ansatz I  (\ref{sol1-GA}) and Ansatz II (\ref{sol2-GA})  with  $\Omega_{k}=0$
we present  one dimensional distributions of $\min\chi^2_\Sigma$ and $\min\chi^2_{tot}$.
For Ansatz I we also draw two dimensional contour plots with notations from
Fig.~\ref{F1}:  the blue lines for $\chi^2_\Sigma$, but the filled contours and red
dashed lines for $\chi^2_{tot}$. We note that the circles and stars in the plots mark minimum points obtained respectively for $\chi^2_{tot}$ and $\chi^2_\Sigma$.}
  \label{F3}
\end{figure*}

We analyze the flat case of Ansatz I (\ref{sol1-GA}) in Fig.~\ref{F3}. For one
dimensional distributions and contour plots we use notations of Fig.~\ref{F1}: the red
dashed lines and filled contours for $\min\chi^2_{tot}$ and the blue lines for
$\min\chi^2_\Sigma$. In 3  panels  of Fig.~\ref{F3} (upper left; upper middle and lower left) we compare this variant of the  model with Ansatz II
(\ref{sol2-GA}) (the green lines). One should note that for both cases the optimal
values of parameters (in particular, for $\lambda_m$, $\lambda_d$) do not coincide for
$\chi^2_{tot}$ and $\chi^2_\Sigma$. In other words, when we include the CMB contribution
$\chi^2_{CMB}$ (\ref{chiCMB}) into the function
$\chi^2_{tot}=\chi^2_{\Sigma}+\chi^2_{CMB}$, the resulting minimum point for
$\chi^2_{tot}$ appears to be shifted. This effect can be seen in the bottom-right panel
of Fig.~\ref{F3}, where for the contour plots  of Ansatz I  the circle and star respectively mark
the minimum points for $\chi^2_{tot}$ and $\chi^2_\Sigma$ (see also the one
dimensional distributions for $\min\chi^2(\lambda_d)$ and $\min\chi^2(\beta)$).

As a consequence of this behavior we have the absolute minima of $\chi^2_{tot}$ for
Ansatz I and Ansatz II in Table~\ref{Estim} only a bit better than the value $576.98$
for the case $w_d={}$const with $\Omega_k=0$. Note that the both variants turn into this
case, if we take $\beta=0$ and $\alpha=0$ in Eqs.~(\ref{sol1-GA}), (\ref{sol2-GA}), respectively.

\begin{figure}
\centerline{\includegraphics[width=0.5\textwidth]{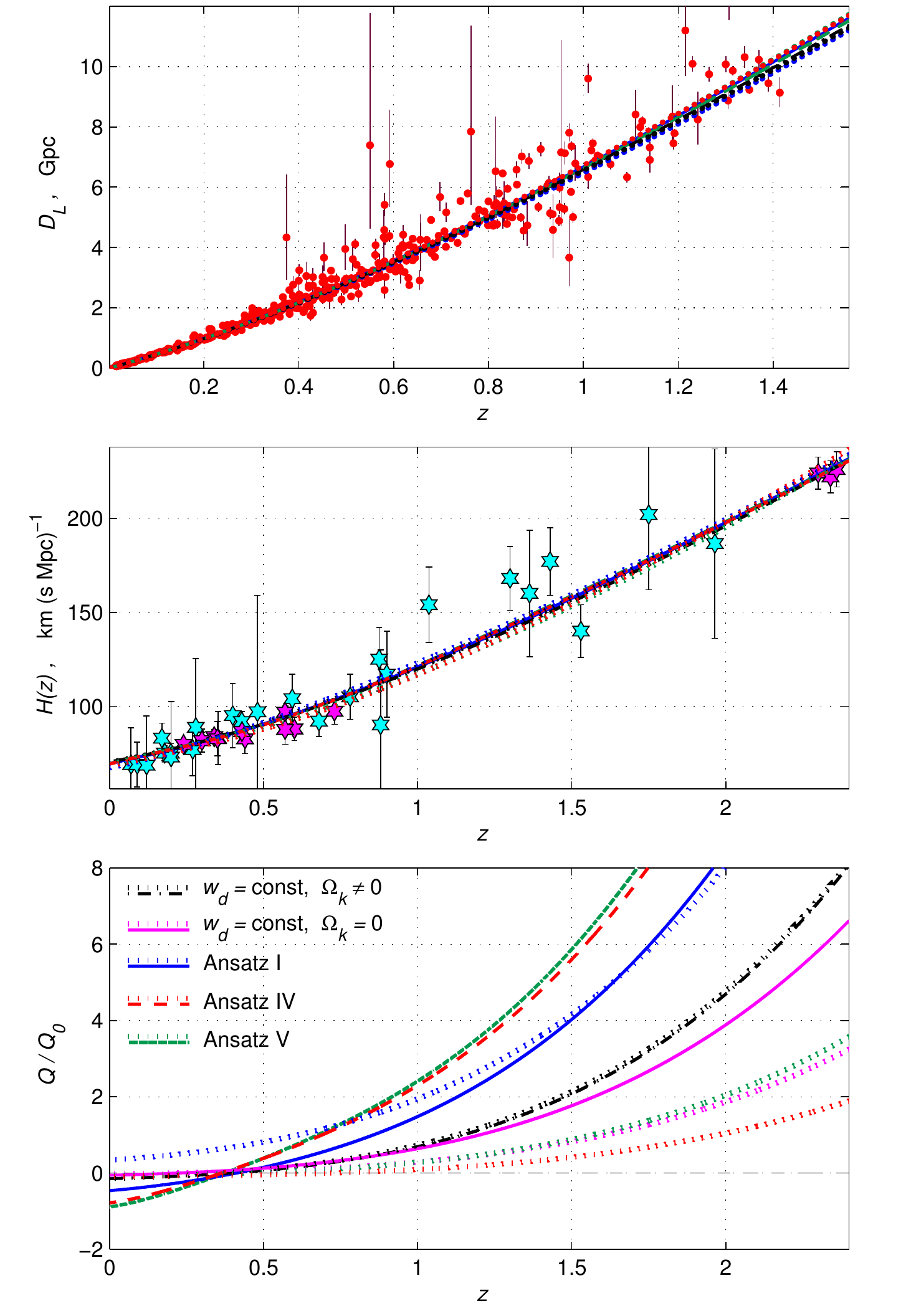}}
 \caption{For different variants of the model with optimal
parameters from Table~\ref{Estim} we show the plots for $D_L(z)$ (upper panel) 
describing the SNe data \cite{Suzuki2012}, $H(z)$ functions with the data from Table~\ref{H-data} (middle panel) and  $Q(z)$ dependence (lower panel). We note that the same labels in the lower panel follow for the 
other two plots (i.e. upper and middle plots). We further note that the plots for different variants of the models both in upper and middle panel are almost indistingushable from each other while although in the lower panel the plots for $Q(z)/Q_0$ are distingushable for large reshift but for low redshifts they are also indistinguishable from each other. }
  \label{FSNHQ}
\end{figure} 
\begin{figure*}
\includegraphics[width=0.95\textwidth]{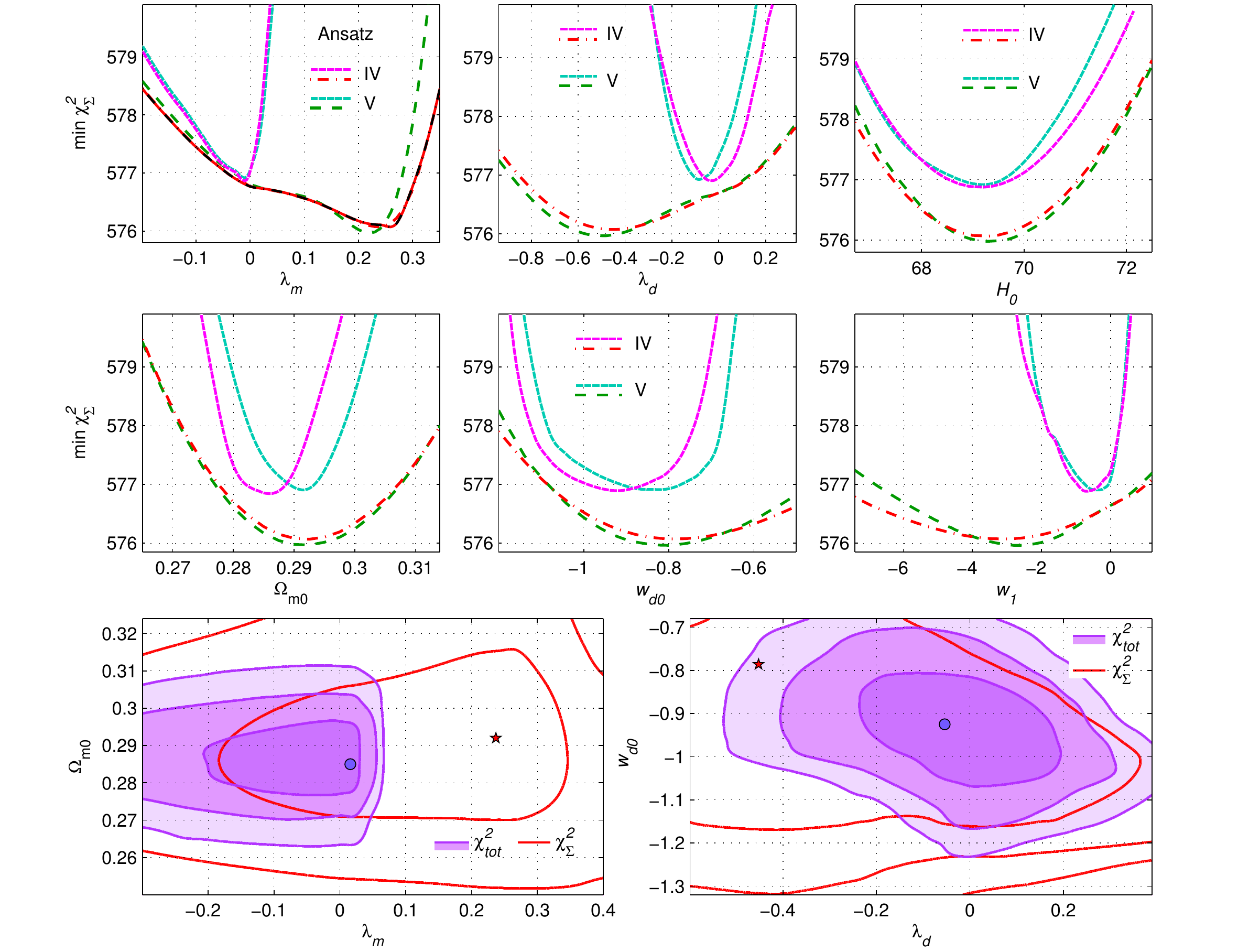}
\caption{Dependence of $\min\chi^2_\Sigma$ on $H_0$, $\lambda_m$, $\lambda_d$,
$\Omega_{m0}$, $w_{d0}$, $w_1$ for  Ansatz IV (\ref{Ans4}) (red and magenta lines) and
for Ansatz V (\ref{Ans5}) (green and aquamarine lines). The contour plots with
$1\sigma$, $2\sigma$ and $3\sigma$ confidence levels in the bottom
panels are shown for Ansatz IV in notations of Fig.~\ref{F1} where the red lines
stands for $\chi^2_\Sigma$ and the filled one for  $\chi^2_{tot}$. The circles and 
stars in the contour plots mark minimum points obtained respectively for $\chi^2_{tot}$ and
$\chi^2_\Sigma$.}
  \label{F4}
\end{figure*} 
When we compare the variants of the model (\ref{GA}), we keep in mind 
that Ansatz I ($\alpha=0$) and Ansatz II (\ref{sol2-GA}) 
($\beta=0$) have the same number of model
parameters $N_p=6$, but  for the case of $w_d={}$const with $\Omega_k=0$, this value is
$N_p=5$. We try to minimize $N_p$, hence, here and below for all variants of the model, we
consider only the flat case ($\Omega_k=0$). In Fig.~\ref{F3} we compare both Ansatz I and Ansatz II where we see that for Ansatz~I, the essential advantage is in the absolute minimum of $\chi^2_\Sigma$, but small for $\chi^2_{tot}$.

Dependence of $\min\chi^2_\Sigma$ and $\min\chi^2_{tot}$ on $\Omega_{m0}$ in the
bottom-left panel of Fig.~\ref{F3} is similar for both presented variants of the model,
but the upper curves for $\chi^2_{tot}$ are more narrow. As usual, these minima are
taken over all remaining parameters. In other panels we see different dependence of
these minima on $\lambda_m$, $\lambda_d$ and  $\beta$. These behavior in some cases are
connected with various singularities (\ref{singul}), which can appear in certain domains
of the parameter space.

In Fig.~\ref{FSNHQ} we demonstrate how the most successful variants of the model
 with parameters from Table~\ref{Estim} describes  SNe data \cite{Suzuki2012} with the functions  $D_L(z)$ (the upper panel) and  $H(z)$ data from
Table~\ref{H-data} (the middle panel); in the bottom panel we draw the corresponding
plots of the interaction function (\ref{interaction1}): $Q(z)= 3H(\lambda_m \rho_{dm}
+\lambda_d \rho_d)$ in its dimensionless form $Q(z)/Q_0$, where
$Q_0=H_0\rho_{cr}=3 H_0^3/(8 \pi G)$.
The  $H(z)$ data in the middle panel are marked as cyan or magenta stars,
if they are obtained from differential ages or from BAO data. From 
Fig.~\ref{FSNHQ}, we see that the plots of $D_L(z)$ and $H(z)$ are practically
coincide for all considered variants.

For the five most successful variants of the model, the dotted lines 
correspond to the optimal parameters for
$\chi^2_{tot}$ (SNe+$H(z)$+BAO+CMB); the solid and dashed lines
describe the minimization of $\chi^2_\Sigma$ (SNe+$H(z)$+BAO). We observe that 
the observational data SNe+$H(z)$+BAO always suggest that there is a transition 
of $Q$ at late time from its positive values to negative values, 
and the transion occurs around $z\simeq 0.4$. On the other hand, for 
the observational data SNe+$H(z)$+BAO+CMB, except for Ansatz I
(in this case $Q$ remains positive throughout the evolution of the universe), 
all other variants keep the same behaviour as we observe for the data 
SNe+$H(z)$+BAO. That means $Q$ changes its sign from positive to negative values
around the same redshift. 
Thus, we find that almost all variants allow the flow of energy
from CDM to DE at late time (precisely for $z \lesssim 0.4$) while at for $z \gtrsim 0.4$, the 
energy flow takes place from DE to CDM. 
Moreover, for $w_d =$ constant (both for $\Omega_k =0$ and $\Omega_k \neq 0$), as seen from the Fig. \ref{FSNHQ}, the quantity $Q/Q_0$ is very very close to zero, that means, a very small interaction 
is favored in this case. It is an interesting result becasue some other interactions also conclude very small interaction in the dark sector for constant $w_d$, see \cite{NPS2016, KN2016, XW2016}.

For the Ansatz III (\ref{Ans3}) the best values of $\min\chi^2_\Sigma$ and $\min\chi^2_{tot}$ 
(see Table~\ref{Estim}) are worse, than the corresponding minima for other
variants of the model with the same $N_p=6$. The main drawback of Ansatz III is 
that its solutions behave badly with the optimal parameters: they are close to singular solutions
of types (a) and (c) (\ref{singul}). So in our calculations of values in
Table~\ref{Estim} we had to bypass singular domains  in the parameter space.
We investigate the same ansatz (Ansatz III)
with the choice $\lambda_m=0$, considered in Ref. \cite{PBC2015} since the background is analytically solved for this case,
see eq. (\ref{variable-total-Energy}).
For this model our analysis shows that $\min\chi^2_\Sigma\simeq576.81$ and
$\min\chi^2_{tot}\simeq577.46$. Thus, it is seen that this variant with analytic solutions
appeared to be unsuccessful in comparison with the case
(\ref{energy-const}): $w_d={}$const, $\Omega_k=0$ (for these variants $N_p=5$, so they
are comparable). Due to these reasons we do not present the constraints for
$\lambda_m=0$, $\alpha=1$ in a separate line in Table~\ref{Estim}. 
Thus, we observe that 
Ansatz III for both choices $\lambda_m \neq 0$ and $\lambda_m = 0$ is not 
suitable as reported by the observational data. So, we do not present 
any graphical analysis for this ansatz.

The next variant (\ref{Ans4}) (Ansatz IV) behaves better for $\chi^2_\Sigma$. 
It has only the type (c1) singularities in the domain
$\lambda_m<0$.  This domain is far from optimal values of the model parameters for
$\chi^2_\Sigma$ with the smallest $\min\chi^2_\Sigma\simeq576.07$ (see Table~\ref{Estim}).
However, the minimum of $\chi^2_{tot}$ is achieved in
the $\lambda_m<0$ domain. So, the minimal value 576.88 of
$\chi^2_{tot}=\chi^2_{\Sigma}+\chi^2_{CMB}$ appears to be rather large.
This behavior is seen in the top-left panel of Fig.~\ref{F4}, where the red dash-dotted
line for $\chi^2_\Sigma$ and the dashed magenta line for $\chi^2_{tot}$ show, how the
minima $\min\limits_{\Omega_{m0},\lambda_d,H_0,w_{d0},w_1}\chi^2$ depend on $\lambda_m$.

For the last variant (\ref{Ans5}) (Ansatz V), 
we achieve the absolute minimum
for $\chi^2_\Sigma$ among all considered models in Tables~\ref{Estim} and \ref{Compar}:
$\min\chi^2_\Sigma\simeq575.97$. 
However, if we add  the CMB, minimum of  $\chi^2_{tot}$ becomes rather large, because it
is achieved near the $\lambda_m<0$ domain. In this domain solutions have the type (c)
singularity (\ref{singul}), so it is practically forbidden.

In Fig.~\ref{F4} one can see the one dimensional distributions of $\min\chi^2_\Sigma$ and
$\min\chi^2_{tot}$ for Ansatz IV  (\ref{Ans4}) and  Ansatz V (\ref{Ans5}). In the bottom
panels we draw two dimensional contour plots for Ansatz IV with filled contours for
$\chi^2_{tot}$ and red lines for $\chi^2_\Sigma$. Here, we use notations from
Figs.~\ref{F1} and \ref{F3}. In particular, the circles and stars demonstrate difference
between minimum points  for $\chi^2_{tot}$ and $\chi^2_\Sigma$. A striking feature that one
must note is in the behaviour of the $w_1$ parameter in Ansatz IV (\ref{Ans4}) and Ansatv V (\ref{Ans5}). From Table \ref{Estim}, one can see that the value of $w_1$ for both 
ansatze (Ansatz IV and Ansatz V) significantly chnages after the inclusion of the CMB data.

\section{Interacting and non-interacting models: A statistical comparison}
\label{comparison}

In this section  we compare our interacting dark energy scenario with some other
existing non-interacting cosmological models purely from the statistical ground. In Table~\ref{Compar} we demonstrate how these models
describe the same observational data for SNe Ia \cite{Suzuki2012}, $H(z)$ and BAO from
Tables~\ref{H-data}, \ref{TBAO}. Calculations were made in accordance with the
procedure described in Sect.~\ref{data-analysis}, and in Ref. \cite{Sharov2016}. Therefore,
we will briefly describe the models in the following subsections.

\subsection{Modified Chaplygin gas and its family}
The equation of state for modified Chaplygin gas (MCG) with pressure $p_g$ and energy
density $\rho_g$  is  \cite{DBC2004,  Benaoum2002}
 \begin{equation}
 p_g= A \rho_g- \frac{B}{\rho_g^{\alpha}}.
\label{MCG} \end{equation}
 Modified Chaplygin gas is the subsequent generalizations of
Chaplygin gas (EoS: $p_g= -B/\rho_g$) and generalized Chaplygin gas (GCG) with EoS
\cite{Kamenshchik2001}:
 \begin{equation}
 p_g= -B/\rho_g^{\alpha}.
 \label{GCG} \end{equation}
In these models GCG or MCG acts as a unified candidate for dark matter and dark energy.

In Table~\ref{Compar} for the MCG and GCG models we cite the results of calculations
from Ref.~\cite{Sharov2016}, where these scenarios were explored as two-component models
with usual dust-like baryonic matter component $\rho_b$ and the Chaplygin gas component
$\rho_g$: $\rho=\rho_b+\rho_g$. In this case the Friedmann 
equation
(\ref{friedmann1}) is
 \begin{eqnarray}
&& H^2/H_0^2= \Omega_{b0} a^{-3}+ \Omega_ka^{-2}\nonumber \\
 &+&(1-\Omega_{b0}-\Omega_k)\Big[B_s+(1-B_s)\,a^{-3(1+A)(1+\alpha)}\Big]^{1/(1+\alpha)}.
  \nonumber\end{eqnarray}
    Here the dimensionless parameter $B_s=B\rho_{g0}^{-1-\alpha}/(1+A)$ is used
instead of $B$, {and} $\rho_{g0}=\rho_g(t_0)$.

For MCG, GCG and other cosmological models  the estimations in Table~\ref{Compar}
were made for the value $\Omega_{b0}=0.044$
(Eq. (\ref{Omb})). It is shown in
Ref. \cite{Sharov2016}, that this value is optimal for the  $\Lambda$CDM,  MCG, GCG and the
model with EoS (\ref{quadr}), if we use the fitting formula (\ref{rsA}). These
estimations were supported with the more simple fitting formula  $r_d=(r_d h)_{fid}\cdot
h^{-1}$, but in the latter case the mentioned models are not sensitive to a value
$\Omega_{b0}$ in the range $0\le\Omega_{b0}\le0.15$, because of similarity in properties
of dark matter and baryonic matter. Due to this reason we do not consider $\Omega_{b0}$
as a free model parameter and fix it in the form (\ref{Omb}) for all models  in
Table~\ref{Compar}.

 One can see
in Table~\ref{Compar} that the MCG model demonstrates the value
$\min\chi^2_\Sigma=576.45$, it is a bit better than the the  $w_d={}$const model
(\ref{energy-const}). The MCG mode also has $N_p=5$ parameters: $H_0$, $\Omega_k$, $A$,
$B_s$, $\alpha$. For the GCG model the minimum of $\chi^2_\Sigma$ is worse, however in
this case we have $N_p=4$ parameters (because $A=0$), so the GCG model gets advantage
from information criteria.

\subsection{Quadratic equation of state}

We consider a cosmic substratum having quadratic equation of state which has similar
unified behavior as in MCG. Further, this quadratic EoS asymptotically becomes of de
Sitter type. The EoS \cite{AB2006, Sharov2016}
 $$
p=\tilde{p}_0+ w_0 \rho_g+ \tilde{\beta} \rho_g^2
 $$
includes the first three terms of the Taylor series expansion of an arbitrary function
$p= f (\rho_g)$, where $\tilde{p}_0$, $w_0$, $\tilde{\beta}$ are free parameters. It is
convenient to rewrite this EoS in the form \cite{Sharov2016}
\begin{equation}
p=p_0 \rho_{cr}+ w_0 \rho_g+ \beta\rho_g^2/\rho_{cr}, \label{quadr}
\end{equation}
where $p_0= \tilde{p}_0/\rho_{cr}$, $\beta= \tilde{\beta}\rho_{cr}$ are the
dimensionless parameters and $\rho_{cr}= 3 H_0^2/ 8 \pi G$.

 \begin{table*}
 {\begin{tabular}{l|c|c|c|c|c|c|c|c}  \hline
 Model   &$ \min\chi^2_\Sigma $ &$ \min\chi^2_{tot} $ &$\dfrac{\min\chi^2_{tot}}{d.o.f} $ & $N_p$ & $AIC_\Sigma$ & $AIC_{tot}$
  & $\Delta\,AIC_ {tot}$ & $\Delta\,BIC_ {tot}$\\ \hline
 $w_d={}$const (non-flat)$\!\!\!$
 & 576.29 & 576.45  & 0.9107 & 6 & 588.29 & 588.45 & 2.46 & 15.84 \\
   \hline
   $w_d={}$const (flat)
 & 576.44 & 576.98  & 0.9101 & 5 & 586.44 & 586.98 & 0.99 & 9.91 \\
  \hline
 Ansatz I [Eq. (\ref{sol1-GA})]
 & 576.29 & 576.93  & 0.9114 & 6 & 588.29 & 588.93 & 2.94 & 16.32 \\
  \hline
   Ansatz IV [Eq. (\ref{Ans4})]
 & 576.07 & 576.90  & 0.9114 & 6 & 588.07 & 588.90 & 2.91 & 16.29 \\
  \hline
  Ansatz V  [Eq. (\ref{Ans5})]
 & 575.97 & 576.92  & 0.9114 & 6 & 587.97 & 588.92 & 2.93 & 16.31 \\
  \hline
 $ \Lambda$CDM$ $
 & 578.56 & 579.99  & 0.9119 & 3 & 584.56 & 585.99 &  0  &  0   \\
  \hline
 GCG [Eq. (\ref{GCG})]
 & 577.01 & 578.26  & 0.9106 & 4 & 585.01 & 586.26 & 0.27 & 4.73 \\
   \hline
 MCG [Eq. (\ref{MCG})]
 & 576.45 & 577.62  & 0.9111 & 5 & 586.45 & 587.62 & 1.63 & 10.55 \\
   \hline
Quadratic [Eq. (\ref{quadr})]
 & 576.03 & 577.46  & 0.9108 & 5 & 586.03 & 587.46 & 1.47 & 10.39 \\
   \hline
CPL [Eq. (\ref{CPL})]
 & 576.57 & 577.83  & 0.9114 & 5 & 586.57 & 587.83 & 1.84 & 10.76 \\
    \hline
Linear [Eq. (\ref{linear})]
 & 576.57 & 577.74  & 0.9113 & 5 & 586.57 & 587.74 & 1.75 & 10.67 \\
 \hline \end{tabular}
 \caption{A statistical comparison of some successful variants of the interacting dark energy model with some well known non-interacting cosmological models has been presented using two different combined analysis SNe+$H(z)$+BAO and SNe+$H(z)$+BAO+CMB.}
\label{Compar}}
 \end{table*}

Solving the conservation equation  (\ref{continuity}), we obtain the energy density
$\rho_g$ in the form
 $$\frac{\rho_g}{\rho_c}=\left\{\begin{array}{ll} \frac{1}{2 \beta}
 \left[\frac{\Gamma-\sqrt{|\Delta|}\tan\left(\frac{x}{2} \sqrt{|\Delta|}\right)}
 {1-|\Delta|^{-\frac{1}{2}}\tan\left(\frac{x}{2} \sqrt{|\Delta|}\right)}-1-w_0\right], &\Delta < 0,\\
 \frac{1}{2 \beta} \left[\left(\frac{x}{2}+\frac{1}{\Gamma}\right)^{-1}-1-w_0\right], & \Delta=0,\\
 \frac{\rho_-(\Omega_m-\rho_{+})\,a^{-3\sqrt{\Delta}}-\rho_{+}(\Omega_m-\rho_{-})}
 {(\Omega_m-\rho_{+})\,a^{-3\sqrt{\Delta}}-\Omega_m+ \rho_{-}}, &
 \Delta > 0;
 \end{array}\right.$$
where $\Delta= (1+w_0)^2- 4 \beta p_0$, $ \Omega_{m}= 1- \Omega_k- \Omega_{b0}$,
$\Gamma= 2 \beta \Omega_m+ 1+ w_0$, $\rho_{\pm}= \frac{-1-w_0\pm
\sqrt{\Delta}}{2\beta}$.

Hence, the evolution equation can be written as
 $$
H^2= H_0^2 \left[\frac{\rho_g}{\rho_c}+ \Omega_{b0} a^{-3}+\Omega_k a^{-2}\right].
 $$

The value $\min\chi^2_\Sigma=576.03$ for the model (\ref{quadr}) in Table~\ref{Compar}
is better than for the MCG model, it is close to the best result of Ansatz V
(\ref{Ans5}).

\subsection{CPL parametrization}

We also consider the universe including cold dark matter and a dark energy component
with Chevallier-Polarski-Linder (CPL) parametrization \cite{CP2001, Linder2003}, i.e.
Eq. (\ref{Ans4})
 \begin{equation}
w = w_0+ w_1 \frac{z}{1+z}= w_0+ w_1(1-a),
 \label{CPL}\end{equation}
where $w_0$, $w_1$ are two free parameters. In presence of this dark energy component
(with its present time fraction $\Omega_{X0}=1-\Omega_{m0} - \Omega_{k}$) the evolution
equation is
 $$ 
\frac{H^2}{H_0^2}=\Omega_{m0} a^{-3}+ \Omega_{k} a^{-2}+ \Omega_{X0} a^{-3(1+w_0+w_1)}
e^{3 w_1(a-1)}.  
 $$ 

It is interesting to compare this model with the considered above variant (\ref{Ans4})
(Ansatz IV) of our interacting model with the similar EoS. In other words, the CPL model
(\ref{CPL}) transforms into Ansatz IV (\ref{Ans4}), if we include the interaction term
(\ref{interaction1}) with two model parameters $\lambda_m$, $\lambda_d$ and fix the
curvature parameter $\Omega_{k}=0$.

From Table~\ref{Compar}  we see that the interacting scenario
(\ref{Ans4}) (Ansatz IV) has the essential advantage ($\min\chi^2_\Sigma\simeq576.07$
and $\min\chi^2_{tot}\simeq576.9$) in compared to the non-interacting model (\ref{CPL})
($\min\chi^2_\Sigma\simeq576.57$, $\min\chi^2_{tot}\simeq577.83$).

\subsection{Linear parametrization}

The model with linear parametrization in EoS is similar to the considered above CPL
parametrization (\ref{CPL}), but has the following EoS \cite{Astier2001, CH1999, WA2002}
 \begin{equation}
w = w_0+ w_1z,
  \label{linear}\end{equation}
 where $w_0$, $w_1$ are two free parameters
to be constrained by the observational data.
{The evolution equation for a universe made of cold matter and the dark energy with the above equation of state} is
 $$
\frac{H^2}{H_0^2}=\Omega_{m0} a^{-3}+ \Omega_{k} a^{-2}+ \Omega_{X0} a^{-3(1+w_0-w_1)}
e^{3 w_1 \left( \frac{1-a}{a}\right)}.
 $$

One can see in Table~\ref{Compar} and in Fig.~\ref{F4} that the model (\ref{linear})
behaves very closely to the CPL scenario  (\ref{CPL}), but it demonstrates essentially
worse $\min\chi^2_\Sigma$ than the corresponding interactive model (\ref{Ans5}) (Ansatz V).

Hierarchy of the scenarios in  Table~\ref{Compar} will change, if we take into account
information criteria which use a number $N_p$
 of model parameters (degrees of freedom). In particular, the Akaike and Bayesian information criteria are given by
\cite{Akaike1974, SHL2012, SKK2006}
 $$
 AIC = \min\chi^2_\Sigma +2N_p,\qquad AIC = \min\chi^2_\Sigma +N_p\log N,
 $$
 where $N$ is the number of data points used in the fit. 
These criteria give advantage to the $\Lambda$CDM and other models with minimal $N_p$.

\section{Summary}
\label{conclu}

In the FLRW background {of our Universe} we have considered an interacting scenario
between dark matter and dark energy where both of them obey barotropic equation of
state.
 The interaction is a linear combination of the energy densities of the dark components
in the form
 $Q= 3\, H\, \lambda_m\, \rho_m+ 3\, H\, \lambda_d\, \rho_d$, where
($\lambda_m$, $\lambda_d$) are the coupling parameters describing the strength and
direction of energy flow from their sign (i.e. whether $Q> 0$ or $Q< 0$). Since the EoS in DE could be either constant
or variable, hence we have examined both the possibilities to explore the cosmological
scenarios with the use of current astronomical data. For $w_d=$ constant, the evolution
equations for matter and dark energy take analytic forms. 
For variable $w_d$ we have proposed three ansatze in Eqns.
(\ref{sol1-GA}), (\ref{sol2-GA}), (\ref{Ans3}), which emerge from the generalized ansatz
given in Eq. (\ref{GA}).
In addition to these, we have considered two more variable EoS in DE in the forms of CPL
and linear parametrizations in equations (\ref{Ans4}), (\ref{Ans5}), respectively. Altogether, we have considered 7 variants for the present interacting model for a detailed analysis.

Henceforth, with the introduction of 7 variants of the EoS in dark energy, we
constrained the model parameters using the joint analysis of Union 2.1, 
Hubble parameter measurements, baryon acoustic oscillation data points and
cosmic microwave background shift parameter. 
We used statistical minimization technique for the $\chi^2$ functions
where we consider two different joint analyses (i)
$\chi^2_{\Sigma}=\chi^2_{SN}+ \chi^2_{H}+ \chi^2_{BAO}$,
and (ii) $\chi^2_{tot}=\chi^2_{SN}+ \chi^2_{H}+ \chi^2_{BAO}+ \chi^2_{CMB}$.  
The results of the analyses are presented 
in Table~\ref{Estim}.

We found that for $w_d= $ constant, the curvature parameter $\Omega_k$ plays significant
role in the analysis. We investigated the cases $\Omega_k \neq 0$ and $\Omega_k= 0$. The
difference in the behavior of $\min\chi^2_\Sigma$ and  $\min\chi^2_{tot}$ for both 
the variants  has been presented in the top panels of
Fig. \ref{F1} and in Table~\ref{Estim}. For the case $\Omega_k\ne 0$ of the model with
$w_d= $ constant, the minimal value  of $\chi^2_\Sigma$ is better,
so for this case we presented two dimensional contour
plots at $1\sigma$, $2\sigma$, $3\sigma$ confidence levels in Fig. \ref{F1}.

Further, the possibility of variable EoS in DE has been investigated with  5 different
variants in  Eqs. (\ref{sol1-GA}) $-$ (\ref{Ans5}) for the present interaction.
We found that the variants may experience singularities (\ref{singul}) (see Fig.
\ref{Fsing}) at finite time, and hence, we excluded the domains of the parameters
leading to the singular behavior and analyzed them by the current data sets mentioned
above. In most of the cases we notice that one of the coupling parameters of the
interaction possesses negative sign,
so during the evolution of the universe the present interaction $Q$ changes its sign,
thereby the direction of energy flow changes. This effect has been shown in Fig.~\ref{FSNHQ} (see the bottom one) for some successful variants of the model. This shows that for $z \lesssim 0.4$, almost all successful variants (except Ansatz I) predict the flow of energy from CDM to DE (i.e. $Q< 0$) while for $z \gtrsim 0.4$, the energy flows from DE to CDM (i.e. $Q>0$).

Furthermore, in figures \ref{F3}, \ref{F4}, 
we have presented the graphical variation of the
$\min\chi^2_\Sigma$ and  $\min\chi^2_{tot}$  over the model parameters for the variable
EoS in DE presented in the paper. For Ansatz I and
 Ansatz IV (CPL parametrization), we
have presented the contour plots in the two dimensional plane for several couple of
model parameters at 1$\sigma$, 2$\sigma$, 3$\sigma$ confidence levels.

For all variants of the model with variable $w_d$, the positions of the minimum points in
parameter spaces are essentially different for the functions $\chi^2_\Sigma$ and
$\chi^2_{tot}$. In fact, we observe that the minimal values of $\chi^2_{tot}$ 
for these variants are larger in these cases. 
Based on the analysis we  may conclude that the interacting model with $w_d= $ constant, 
is the most successful one in respect to all observational data.

Finally, some of the successful variants of the interaction model, 
such as, $w_d= $ constant (with $\Omega_k =0$ and $\Omega_k \neq 0$), Ansatz I,
Ansatz IV, Ansatz V have been compared with some
known non-interacting cosmological models, such as, $\Lambda$CDM model, unified models, namely, the
generalized Chaplygin gas (GCG), modified Chaplygin gas (MCG), a fluid with quadratic
equation of state, and finally with CPL and linear parametrizations in DE. The results
have been presented in Table  \ref{Compar}.

It is found that the present interacting DE model with constant $w_d$  slightly favors
the phantom region in agreement with the latest report \cite{NPS2016}. The best
absolute value of $\min\chi^2_\Sigma$ is achieved for the  interacting model with EoS
(\ref{Ans5}) (Ansatz V). The second result demonstrates that among the non-interacting
models the model with quadratic  EoS  in (\ref{quadr}) provides a better fit with the
observational data. However, the number of model parameters of this non-interacting
model ($N_p=5$) is less than the number of model parameters for the interacting model
(\ref{Ans5}), but larger than the number of model parameters in models GCG or
$\Lambda$CDM.
Finally, we notice that although different models have different model parameters, still
from AIC and BIC  analysis, the models presented in Table \ref{Compar} do not deviate so
much from the $\Lambda$CDM model with minimum number of model parameters in comparison
with others.

\section*{ACKNOWLEDGMENTS}
The authors thank the referee for some essential comments to improve the work.
SP was supported by the Science and Engineering Research Board through NPDF (File No:
PDF/2015/000640).


\end{document}